\documentclass[reprint,amsmath,amssymb]{revtex4-1}
\usepackage[utf8x]{inputenc}
\usepackage{graphicx}
\usepackage{dcolumn}
\usepackage{bm}
\usepackage{float}
\usepackage{xcolor}

\begin{document}

\title{Generalized Aubry-Andr{\'e}-Harper model with modulated
   hopping  and $p$-wave pairing}
\author{M. Yahyavi}
\email{m.yahyavi@bilkent.edu.tr}
 \affiliation{Department of Physics, Bilkent University, TR-06800
  Bilkent, Ankara, Turkey}

\author{B. Het\'enyi}
\email{hetenyi@fen.bilkent.edu.tr,
  hetenyi@phy.bme.hu}
 \affiliation{Department of Physics, Bilkent University, TR-06800
	Bilkent, Ankara, Turkey}
\affiliation{Department of
	Theoretical Physics and MTA-BME ``Momentum'' Topology and
	Correlation Research Group, Budapest University of Technology and
	Economics, 1521 Budapest, Hungary}

\author{B. Tanatar}
\email{tanatar@fen.bilkent.edu.tr}
\affiliation{Department of Physics, Bilkent University,  TR-06800
	Bilkent, Ankara, Turkey }

\begin{abstract}
We study an extended Aubry-Andr{\'e}-Harper model with simultaneous
modulation of hopping, on-site potential, and $p$-wave superconducting
pairing.  For the case of commensurate modulation of $\beta = 1/2$ it
is shown that the model hosts four different types of topological
states: adiabatic cycles can be defined which pump particles, two
types of Majorana fermions, or Cooper pairs.  In the incommensurate
case we calculate the phase diagram of the model in several regions.
We characterize the phases by calculating the mean inverse
participation ratio and perform multi-fractal analysis.  In addition,
we characterize whether the phases found are topologically trivial or
not.  We find an interesting critical extended phase when
incommensurate hopping modulation is present.  The rise between the
inverse participation ratio in regions separating localized and
extended states is gradual, rather than sharp.  When, in addition, the
on-site potential modulation is incommensurate, we find several sharp
rises and falls in the inverse participation ratio.  In these two
cases all different phases exhibit topological edge states.  For the
commensurate case we calculate the evolution of the Hofstadter
butterfly and the band Chern numbers upon variation of the pairing
parameter for zero and finite on-site potential.  For zero on-site
potential the butterflies are triangular-like near zero pairing, when
gap-closure occurs, they are square-like, and hexagonal-like for
larger pairing, but with the Chern numbers switched compared to the
triangular case.  For the finite case gaps at quarter and
three-quarters filling close and lead to a switch in Chern numbers.
\end{abstract}

\maketitle

\section{Introduction}
\label{sec:intro}

The physics of Anderson delocalization-localization (or
metal-insulator) transition in disordered fermionic systems is a
problem of long-standing interest in condensed matter
physics~\cite{Anderson58,Evers08,Abrahams2010}.  In one dimension
uncorrelated random potentials lead to complete localization of all
eigenfunctions~\cite{Shima04,Izrailev12}, while mobility edges and the
delocalization-localization transition will typically appear in 3D
disordered systems.  However, mobility edges may occur in some 1D
systems \cite{Biddle11,Zhang15arXiv,Wang17}, if the disorder
distribution is deterministic, rather than uncorrelated.  The paradigm
of this class of quasi-periodic systems, incommensurate lattices (the
superposition of two periodic lattices with incommensurate periods) is
the Aubry-Andr{\'e} model~\cite{AA80}, or its two-dimensional analog,
the Harper~\cite{Harper55} model. The delocalization-localization
transition due to the disordered on-site potential can appear in the
Aubry-Andr{\'e} model when the lattice is incommensurate, arising from
the self-duality of this model~\cite{Aulbach04}.

The model was explored~\cite{AA-Top1,AA-Top2,AA-Top3} from a
topological perspective.  In the Aubry-Andr{\'e} model with $p$-wave
superconducting (SC) pairing the connection between the
Su–Schrieffer–Heeger-like~\cite{Su79} and the Kitaev-like
\cite{Wakatsuki14,Kitaev01} topological phases was
investigated~\cite{Zeng16}.  Other
studies~\cite{Thouless88,Sarma88,Sarma10,Liu18} focused on
localization effects. In addition, commensurate and incommensurate
modulations may appear in on-site and hopping terms, the interplay
between the two was studied in Ref. ~\onlinecite{Cestari16}.  When
hopping modulations are incommensurate,~\cite{Liu15} the system will
go through Anderson-like localization, but no mobility edge is found.
For commensurate hopping modulations topological zero-energy edge
modes are found~\cite{Liu15,Ganeshan13}.  In
Ref. ~\onlinecite{Cestari16} the incommensurate and commensurate
off-diagonal modulations were combined resulting in the conclusion
that the states depend on the phase between the two.  Experimentally
the model was realized in ultracold atoms in optical
lattices~\cite{Chabe,Roati08} and in photonic
crystals~\cite{Negro03,Lahini09}.  A recent experiment~\cite{Kraus12},
realized the topological edge state.

In this paper, we study a generalized AAH model with modulated on-site
potential, hopping, and $p$-wave pairing.  For the bi-partite case
($\beta = 1/2$) we show that four different topological excitations
are possible.  \textcolor{red}{The same model, but without modulation
  of the $p$-wave pairing was studied by Zeng et al.~\cite{Zeng16} and
  Liu et al.~\cite{Liu18}.  They studied both the commensurate and
  incommensurate cases.  They mapped the phase diagram of the model,
  studied localization by investigating the mean inverse participation
  ratio (MIPR), and did multi-fractal analysis in the incommensurate
  case and showed the existence of topological edge states in the
  commensurate case.  We also do these calculations for the model with
  modulated $p$-wave pairing.  The MIPR studies of the critical
  extended phases give an interesting result.  When incommensurate
  hopping modulation is turned on we see a ``smeared mobility edge''
  phase, in which the rise in the MIPR between the localized and
  extended regions is gradual, rather than sharp.
  (Fig. \ref{fig:IPR_fig5}).  Other GAAHs all show a sharp
  jump~\cite{Sarma10,Liu15} in mobility edge phases.  When, in
  addition, incommensurate on-site potential modulation is turned on,
  the rise in MIPR between localized and extended regions are sharp
  again, but there are more than one such jumps in MIPR.  In these
  last two incommensurate studies topological edge states exist in all
  phases.}  Also, for the commensurate lattice, we investigate the
Chern numbers of the main gaps during the change of the modulated
$p$-wave SC pairing strength.  We find that values of the Chern
numbers are changing with and without on-site potential when we tune
the modulated $p$-wave SC pairing strength.  The modulated $p$-wave
pairing strength changes, the energy spectrum alters from the
triangular-lattice like Hofstadter butterfly to one which is
square-lattice like.

Our paper is organized as follows: In the next section
(Sec. \ref{Sec2}), the generalized version of the GAAH model that
includes nearest-neighbor and next-nearest-neighbor $p$-wave SC
pairing is defined on the infinite lattice. In Sec. \ref{Sec3}, we
extend the 1D model to an ``ancestor" 2D $p$-wave SC model. In
Sec. \ref{Sec_CBhalf}, we check topological properties of the pure
commensurate lattice for the $\beta = {1/2}$. In Sec. \ref{Sec_IC}, we
consider the incommensurate modulations case for $\beta$, where we
will discuss the metal-insulator transition, and especially the
influences of the modulated $p$-wave SC pairing strength on this
transition. In Sec. \ref{Sec_C}, for pure commensurate lattice, the
corresponding Hofstadter butterflies are discussed in detail. We
conclude the paper in Sec. \ref{Con}.

\begin{figure}[ht]
\includegraphics[width=1.1\linewidth]{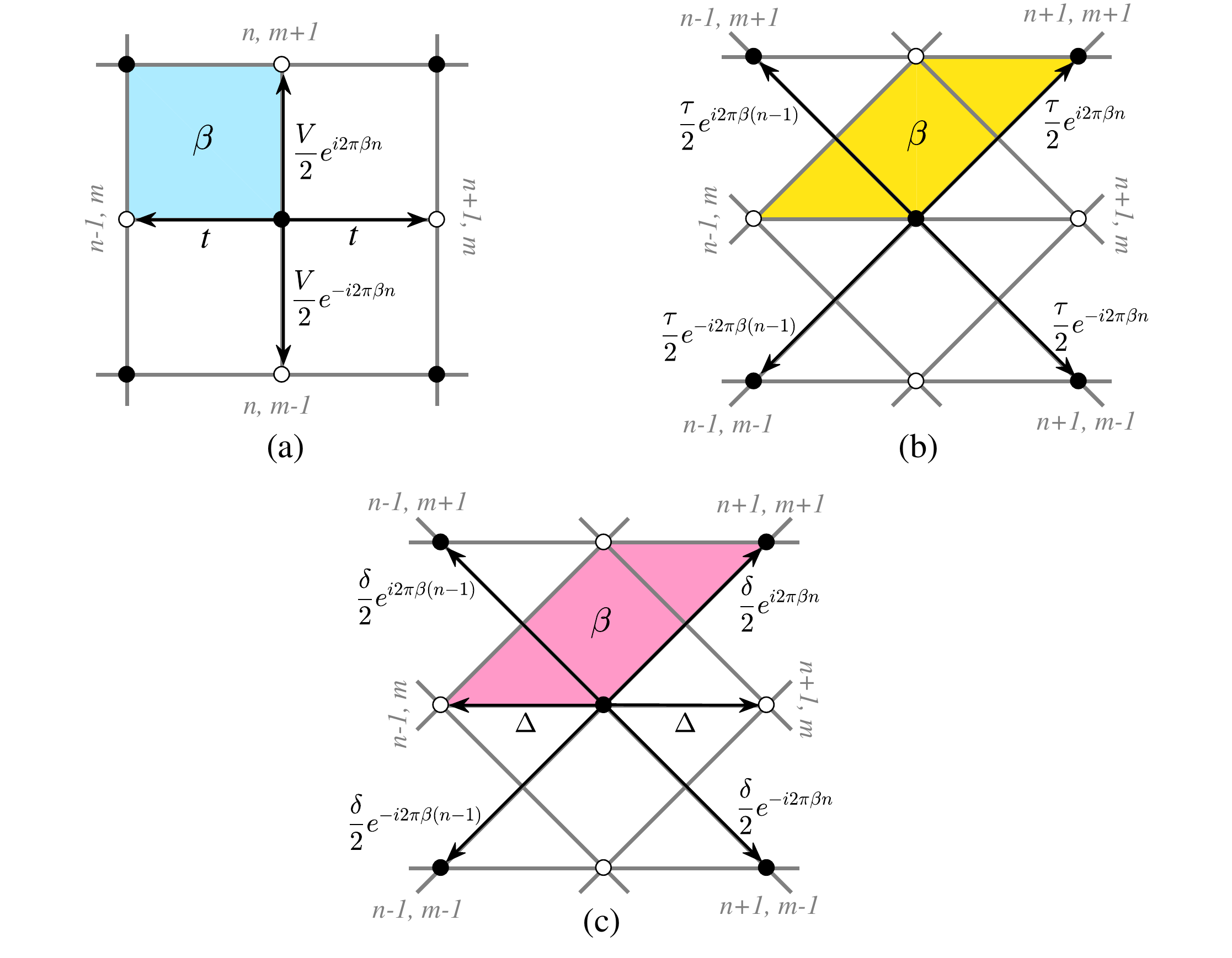}
\caption{Graphical presentations of extended 1d GAAH with nearest and
  next-nearest hopping (SC pairing) and on-site potential to a 2D
  Hamiltonian.  In the presence of a perpendicular magnetic field with
  $\beta$ flux quanta per unit cell, the electrons hop on a
  rectangular lattice. (a) This is the 2D ``ancestor" of the diagonal
  GAAH model which the hopping is to nearest neighbors, (b) This is
  the 2D ``ancestor" of the off-diagonal GAAH model which the hopping
  is to next-nearest neighbors, and (c) This is the 2D ``ancestor" of
  the SC pairing of GAAH model which the hopping is to nearest and
  next nearest neighbors.  Each rectangular plaquettes shown in
  different colors are pierced by $\beta$ flux
  quanta. }\label{AA-DWt0}
\end{figure}

\section{The Model}\label{Sec2}
The generalized one-dimensional Aubry-Andr{\'e}-Harper model with
$p$-wave SC pairing which we study here is described by the following
Hamiltonian
\begin{equation} \label{GAA-PW}
\begin{split}
\hat{{H}}&=-\sum_{j} \left[(t+\tau_j)c_{j}^\dag c_{j+1} + \mbox{H.c.})\right] \\
  &\;\;\;\; +\sum_{j}(\Delta+\delta_j) c_{j}^\dag c_{j+1}^\dag + \mbox{H.c.})+\sum_{j} V_j \hat{n}_j
\end{split}
\end{equation}
where
\begin{equation}
\begin{cases}
     \tau_j=\tau \cos(2\pi \beta j+\phi_{\tau}),  \\    
     \delta_j=\delta \cos(2\pi \gamma j+\phi_{\delta}),  \\     
     V_j=V \cos(2\pi \beta j+\phi_{V}).              
\end{cases}
\end{equation}
Here $\tau_j$ is commensurate (incommensurate) hopping modulations
with periodicity $1/\beta$ and phase factor $\phi_{\tau}$, and $V_j$
is the diagonal Aubry-Andr{\'e} potential with periodicity $1/\beta$
and phase factor $\phi_{V}$, respectively. The corresponding hopping
modulation amplitude is set by $\tau$, and $V$ is the on-site
potential strength.  $\hat{n}_j =c_{j}^\dag c_{j}$ are number
operators, $c_{j}^\dag$($c_{j}$) are creation (annihilation operators)
at position $j$ on the lattice, and $t$ is the hopping (or tunneling)
amplitudes to the nearest neighbors and set to be the unit of the
energy ($t=1$). Also, $\delta_j$ is a SC modulation with periodicity
$1/\beta$ and phase factor $\phi_{\delta}$. Here $\delta$ and $\Delta$
are the strengths of SC pairing gap taken to be real.

In the limit $\Delta=\delta=0$, this model reduces to the generalized
Aubry-Andr{\'e}-Harper model introduced by Ganeshan \textit{et
  al.}~\citep{Ganeshan13}. If $\delta= 0$ and $\phi_{\tau}=\phi_{V}$,
this model reduces to the GAAH model with $p$-wave SC pairing
introduced in Ref. \cite{Zeng16} and studied in Ref. ~\cite{Liu18}. If
we set $\tau=\delta=0$ the model exhibits an Anderson localization
transition when $V > 2(t + \Delta)$ \cite{Cai13}. On the other hand,
if $\Delta$, and $\delta$ are zero, but $\tau$ and $V$ are finite, and
when the relation between the hopping modulation and on-site phases
are fixed, for example, $\phi_V = \phi_\tau + \beta \pi$, the GAA
model can be formally derived from an ancestor 2D quantum Hall system
on a lattice (Hofstadter model) with diagonal (next-nearest-neighbor)
hopping terms~\cite{Hiramoto89,Han94,Kraus12}. Here we keep our
notations general with $\phi_{\tau}=\phi_{\delta}=k_y$ and
$\phi_{V}=k_y+\varphi$ as independent variables. The off-diagonal
modulation has an additional phase $\varphi$.  

\begin{figure}[ht]
    \centering
    \includegraphics[width=\columnwidth]{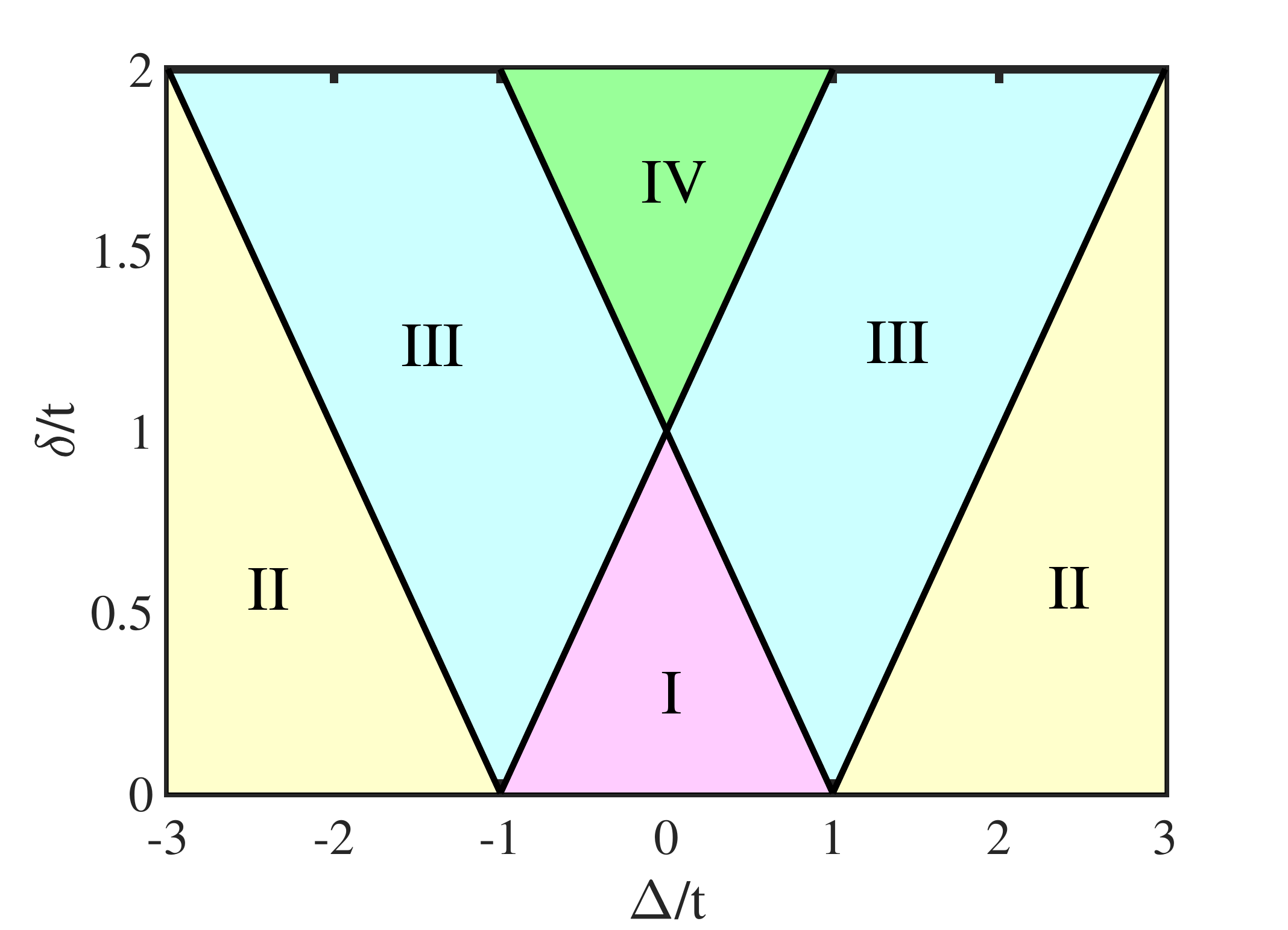}
    \caption{ Phase diagram of the off-diagonal GAAH model as a
      function of the $p$-wave incommensurate modulation amplitude
      $\delta$ and the $p$-wave pairing strength $\Delta$. The hopping
      incommensurate modulation amplitude is set to $\tau = 0$ and the
      phase in the incommensurate modulation is set to $k_y = \pi/2$
      and $\varphi= \beta \pi$. The phases are (I and II) extended
      phases, (III) topological critical phase, and (IV)
      non-topological critical phase.  }
     \label{fig:PDv0}
\end{figure}

\begin{figure}
	\centering
	\includegraphics[width=\columnwidth]{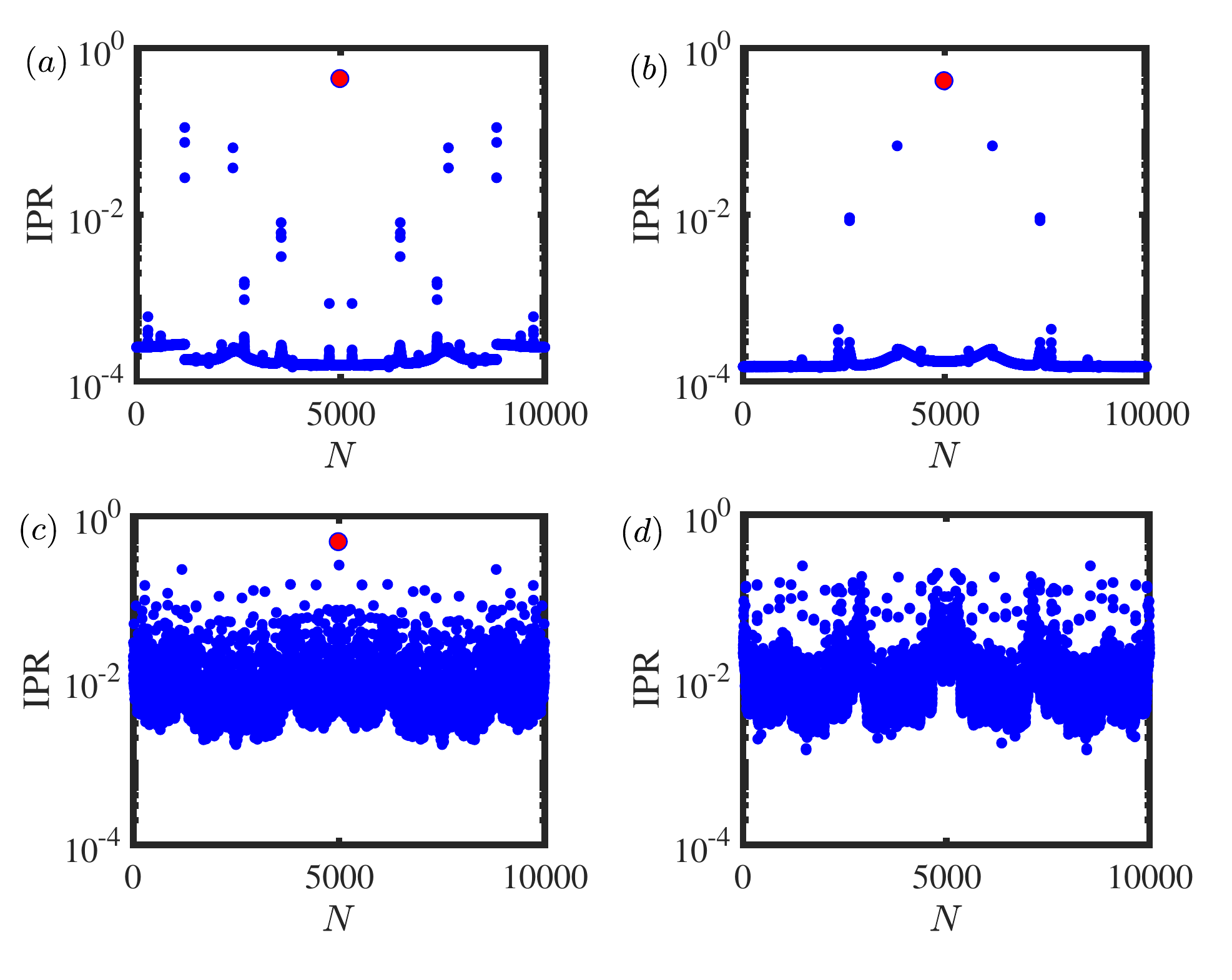}
	\caption{(Color online) The distribution of IPRs over all the
          eigenstates for (a) and (b) extended phases, (c) topological
          critical phase, and (d) non-topological critical phase of
          Fig. \ref{fig:PDv0}. } \label{fig:IPR_fig2}
	\label{fig:muf2}
\end{figure}

\begin{figure}[ht]
    \centering
    \includegraphics[width=\columnwidth]{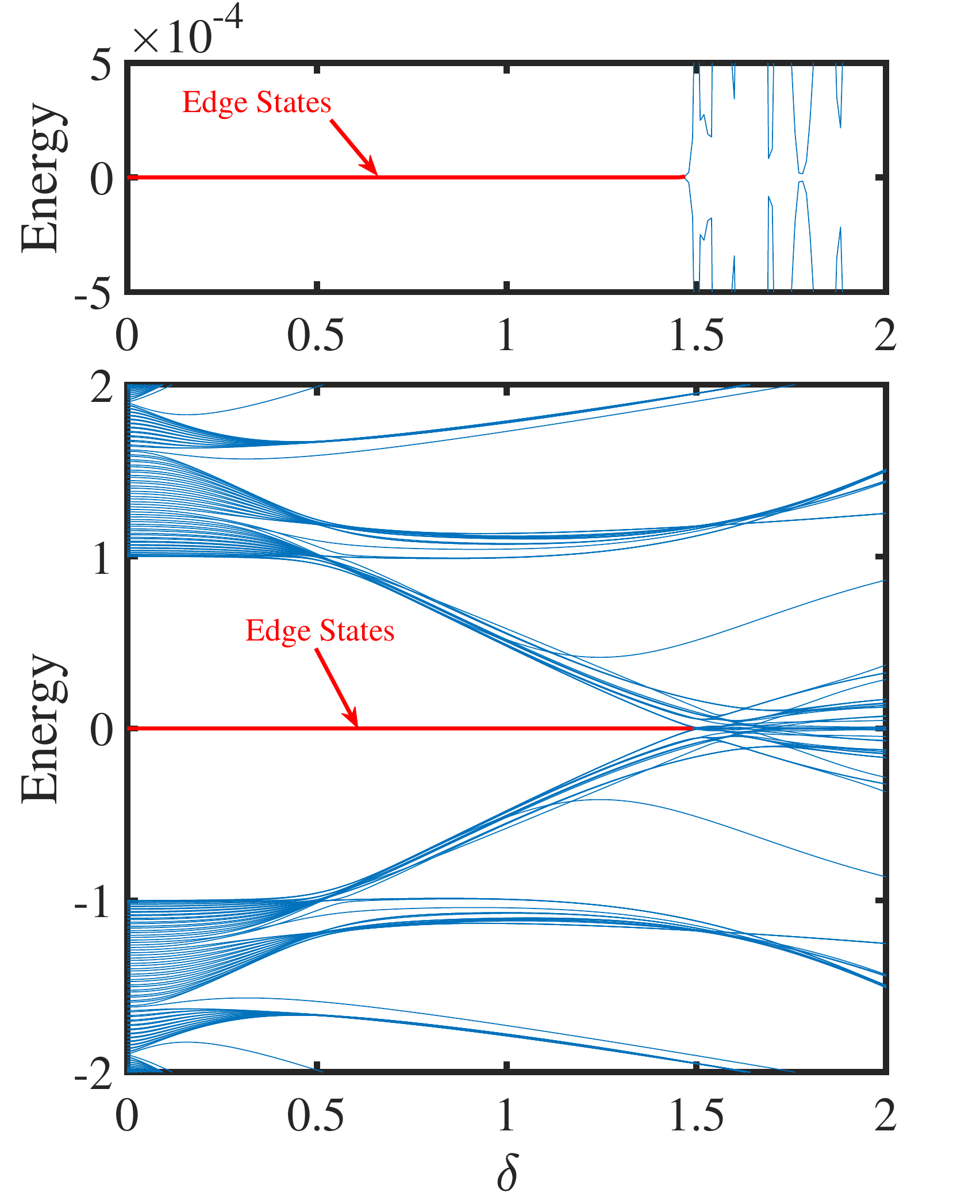}
    \caption{The energy spectrum of the off-diagonal GAAH model with
      $p$-wave pairing plotted as a function of $\delta$ under OBCs
      with $L=100$ lattice sites. The model parameters are $\tau=0$,
      $V=0$, and $\Delta=0.5$.  Inside the regions I and II (see
      figure \ref{fig:PDv0}) there are edge states.  }
     \label{fig:EOBC1}
\end{figure}

 \section{The 2D analog of the generalized Aubry-Andr\'e model with $p$-wave superconductivity}

 \label{Sec3}
 
The Hamiltonian of the 1D GAAH model with $p$-wave superfluid pairing
we consider in this paper can be made to correspond to a 2D $p$-wave
SC model.  For any given $k_y$, the GAAH model of Eq. (\ref{GAA-PW})
can be viewed as the $k_y$th Fourier component of a general 2D
Hamiltonian. On the other hand, $k_y$ is the second degree of freedom,
hence, we define the operator $c_{n,k_y}$ that satisfies the following
commutation relation
\begin{equation}
	\{c_{n,k_y},c^\dagger_{\acute{n},\acute{k_y}}\}=\delta_{n,\acute{n}}\delta_{k_y,\acute{k_y}}
\end{equation}
Therefore the 2D Hamiltonian can be expressed in terms of $\hat{H}$ as
\begin{equation} \label{eqn:H_AA}
  \hat{\mathcal{H}} =\frac{1}{2\pi}\int_{0}^{2\pi} \hat{H}(k_y) \;dk_y
\end{equation}
where in the Hamiltonian of Eq. (\ref{GAA-PW}), we replaced the
operators $c_n$ with $c_{n,k_y}$. The corresponding Hamiltonian can be
written as
\begin{equation}
  \label{eqn:H_AA}
  \begin{split}
    \hat{H}(k_y)&=-\sum_{n}\left[(t+\tau_n)c_{n,k_y}^\dag c_{n+1,k_y}
      + \mbox{H.c.}\right]
    \\ &\;\;\;\;+\sum_{n}\left[(\Delta+\delta_n) c_{n,k_y}^\dag
      c_{n+1,-k_y}^\dag + \mbox{H.c.})\right] \\&\;\;\;\;+\sum_{n} V_n
    c_{n,k_y}^\dag c_{n,k_y}
  \end{split}
\end{equation}
Fourier transforming only in the $y$-direction
\begin{equation}
  \label{eqn:H_AA}
  c_{n,k_y}=\sum_{m} e^{-ik_y m} c_{n,m}
\end{equation}
allows us to easily calculate the  2D Hamiltonian as
\begin{equation}
  \label{eqn:HH_AA2}
\begin{split}
	\hat{\mathcal{H}}&=\sum_{n,m}-\left\{ t c_{n,m}^\dag c_{n+1,m} \right. \\
	& -\left[\frac{\tau}{2} \left((e^{i2\pi \beta n} c_{n,m}^\dag  c_{n+1,m+1}+ e^{-i2\pi \beta n}    c_{n,m}^\dag  c_{n+1,m-1}\right)\right]\\
	& +\left[\frac{\delta}{2}\left(e^{i2\pi \beta n} c_{n,m}^\dag  c^\dag_{n+1,m+1}+e^{-i2\pi \beta n}  c_{n,m}^\dag  c^\dag_{n+1,m-1}\right)\right.\\
	&  \left.+\Delta c_{n,m} c_{n+1,m} \right] 
	+ \frac{V}{2}e^{i(2\pi \beta n+\varphi)} c_{n,m}^\dag  c_{n,m+1} \left.+ \mbox{H.c.}\right\}.
\end{split}
\end{equation}
When $\varphi=0$, the 2D system has isotropic next-nearest-neighbor
hoppings and the corresponding 1D Hamiltonian has the same phase $k_y$
in the off-diagonal and diagonal modulations. This Hamiltonian
describes a 2D lattice in the presence of a uniform perpendicular
magnetic field with $\beta$ flux quanta per unit cell as shown in
Fig. \ref{GAA-PW}.

\section{Commensurate modulation, the case of $\beta = \frac{1}{2}$}
\label{Sec_CBhalf}
Setting $\beta = \frac{1}{2}$ in Eq. (\ref{GAA-PW}), and all three
phases to zero, the lattice becomes bi-partite.  Introducing the
notation $c_i$ and $d_i$ for the two sublattices, Fourier transforming
using the Nambu basis ${c_k^\dagger,d_k^\dagger,c_{-k},d_{-k}}$, the
Hamiltonian becomes
\begin{eqnarray}
  \label{Hbeta_half}
  \hat{{H}} = - 2 t \cos(k) \sigma_z \otimes \sigma_x
  - 2 \tau \sin(k) \sigma_z \otimes \sigma_y  \hspace{.1in} \\ \nonumber + V \sigma_z \otimes \sigma_z 
  - 2 \Delta \sin(k) \sigma_y \otimes \sigma_x
  + 2 \delta \cos(k) \sigma_y \otimes \sigma_y.
\end{eqnarray}
Note that the first three terms correspond to $\sigma_z \otimes
\hat{{H}}_{RM}$, where $\hat{{H}}_{RM}$ is the Rice-Mele Hamiltonian.
It follows that if we set $\Delta = \delta = V = 0$ the model consists
of two independent SSH models (which are topological), and it is
possible to define an adiabatic process in which charge is pumped
across the unit cell (the topological edge states support charges
localized at the edges of the system).  In a similar vein, keeping
$V=0$, it is possible to pair the two $\cos(k)$ with the two $\sin(k)$
terms in four ways.  Keeping the other parameters zero leads to other
possible topological states, or possible adiabatic pumping processes.
Of the remaining three, two are Majorana fermions, and the fourth one
a Cooper pair.  For example, we can take $\tau = \delta = 0$ resulting
in
\begin{equation}
  \hat{{H}} = [- 2 t \cos(k) \sigma_z - 2 \Delta \sin(k) \sigma_y ]
  \otimes \sigma_x,
\end{equation}
in other words, a $2\times 2$ Hamiltonian of the SSH form, but the
adiabatic pumping in this case would not correspond to a charge pump,
because different members of the Nambu bases are coupled by the matrix
elements.  For each of the four cases it is possible to construct the
time-reversal, particle-hole symmetry operators, as well as the chiral
symmetry operators using the ``left'' or the ``right'' part of the
direct product and multiplying with an identity operator from the
other side.  They all fall in the BDI symmetry class~\cite{Altland97}.

\section{Incommensurate modulation}\label{Sec_IC}

When $\beta$ is irrational, the lattice is incommensurate. We choose
$\beta= (\sqrt{5}-1)/2$, the golden ratio, but all the conclusions can
also be generalized to other incommensurate situations.  In this
paper, for incommensurate modulation case, we shall study the
interplay of the SC modulation pairing $\delta$ with the
incommensurate hopping amplitude and potential, respectively, and then
we determine the phase diagram of the model. The Hamiltonian can be
diagonalized by the Bogoliubov–de Gennes (BdG)
transformation~\cite{Lieb61,Gennes66}
\begin{equation}
\eta_n^\dagger=\sum_{j=1}^{L}[u_{n,j}c_{j}^\dag+ \nu_{n,j}c_{j} ]
\end{equation}
where $n=1,\ldots L$, is the energy band index and $u_{n,j}$ and $\nu_{n,j}$ denote the two wavefunction components at the site $j$ assumed to be real. On this basis the wave function of the Hamiltonian becomes
\begin{equation}\label{BGWF1}
|\Psi_n \rangle=\eta_n^\dagger |0 \rangle=\sum_{j=1}^{L}[u_{n,j}c_{j}^\dag+ \nu_{n,j}c_{j} ] |0 \rangle
\end{equation}
Then the Hamiltonian in Eq. (1)  can be diagonalized in terms of the operators $\eta_n$ and $\eta_n^\dagger$ as
\begin{equation}
\hat{H}=\sum_{n=1}^{L} \varepsilon_n (\eta_n^\dagger \eta_n-\frac{1}{2})
\end{equation}
with $\varepsilon_n$ being the spectrum of the quasiparticles. The Schr{\"o}dinger equation $H|\Psi_n \rangle=\varepsilon_n |\Psi_n \rangle$, can be written as
 \begin{widetext}
\begin{equation}
\begin{cases}
-(t+\tau_{j-1})u_{n,j-1}+(\Delta+\delta_{j-1}) \nu_{n,j-1}+V_j u_{n,j}-(t+\tau_j)u_{n,j+1}-(\Delta+\delta_{j}) \nu_{n,j+1} &=\varepsilon_n u_{n,j}  \\
\;\;\; (t+\tau_{j-1})\nu_{n,j-1}-(\Delta+\delta_{j-1}) u_{n,j-1}-V_j \nu_{n,j}+(t+\tau_j)\nu_{n,j+1}+(\Delta+\delta_j) u_{n,j+1}&=\varepsilon_n \nu_{n,j} .
\end{cases}
\end{equation}
\end{widetext}
Representing the wave function as
\begin{equation}
|\Psi_n \rangle=[u_{n,1},\nu_{n,1},u_{n,2},\nu_{n,2},\cdots,u_{n,L},\nu_{n,L}]^T,
\end{equation}
the  Hamiltonian $H$ can be written as a $2L \times 2L$ matrix,
\newpage
\begin{equation}
H_n=\begin{pmatrix}
 A_1 & B & 0& \cdots & \cdots & \cdots & C \\
  B^\dagger & A_2 & B & 0&\cdots & \cdots & 0 \\
   0 &B^\dagger & A_3 & B &  \cdots & \cdots & 0 \\
  \vdots & \ddots & \ddots & \ddots & \ddots & \ddots & \vdots \\
  0 & \cdots & 0&B^\dagger & A_{L-2} & B & 0\\
  0 & \cdots & \cdots &0 & B^\dagger & A_{L-1} & B \\
  C^\dagger &  \cdots & \cdots & \cdots &0 & B^\dagger & A_L
 \end{pmatrix}
\end{equation}
where
\begin{equation}
A_j=\begin{pmatrix}
 V_j & 0 \\
  0 &  -V_j
 \end{pmatrix},
\end{equation}
\begin{equation}
B=\begin{pmatrix}
 -(t+\tau_j) & -(\Delta+\delta_{j}) \\
  \Delta+\delta_{j} &  t+\tau_j
 \end{pmatrix},
\end{equation}
 and  
\begin{equation}
 C=\begin{pmatrix}
 -(t+\tau_{j+1})  &  \Delta+\delta_{j+1} \\
  -(\Delta+\delta_{j+1}) &  t+\tau_{j+1}
 \end{pmatrix},
\end{equation}
for the lattice with periodic boundary conditions, or 
\begin{equation}
C=\begin{pmatrix}
0  &  0\\
0 &  0
\end{pmatrix},
\end{equation}
for the lattice with open boundary conditions. Here we consider, a
chain of length $L$ with periodic boundary conditions.  The irrational
$\beta$ can be approximated by a sequence of rational
numbers~\cite{Wang16} (see Eq. (\ref{Fn})).  In our model we may expect the
usual~\cite{Aulbach04} delocalization transition that occurs in the
original AA model in the incommensurate case.  To show this, we
calculate the mean inverse participation ratio (MIPR) which for a
given normalized wave function ($\sum_{j=1}^{N} (u_{n,j}^2 +
\nu_{n,j}^2)=1$) defined as~\cite{Thouless72,Kohmoto83}
\begin{equation}\label{MIPR_Eq}
  \textmd{MIPR} = \frac{1}{2N}\sum_{n=1}^{2N}\sum_{j=1}^{N} (u_{n,j}^4 + \nu_{n,j}^4)
\end{equation}
where $n$ is the index of energy levels and $u_{n,j}$ and $\nu_{n,j}$
are the solution to the BdG equations. It is well known that for an
extended state, MIPR$\rightarrow \frac{1}{L}$ and the MIPR tends to
zero in the thermodynamic limit (for large $L$), however, MIPR tends
to a finite value for a localized state even in the thermodynamic
limit. In the following, we will calculate the MIPR for different
configurations of our GAAH model with $p$-wave pairing for generic and
off-diagonal cases to characterize the phase boundaries separating 
localized, critical, and extended phases.

Next, in order to clarify the nature of different phases in our model,
we perform multifractal analysis~\cite{Hiramoto89} of the
eigenfunctions, a technique which was applied to study the
quasiperiodic chain with $p$-wave pairing~\cite{Liu18} and also the
original Aubry-Andr{\'e} model~\cite{Hiramoto89,Kohmoto08}. From the
above assumption regarding the irrational value of $\beta$, the golden
ratio can be approached by the Fibonacci numbers via the relation
\begin{equation}\label{Fn}
\beta =\lim\limits_{m \rightarrow\infty} \frac{F_{m-1}}{F_m},
\end{equation}
where $F_m$ is the $m$-th Fibonacci number.  We choose the chain $L =
F_m$. It is recursively defined by the relation $F_{m+1} = F_m +
F_{m-1}$, with $F_0 = F_1 = 1$. The probability measure can be defined
from a wave function of Eq. (\ref{BGWF1}) as
\begin{equation} 
p_{n,j} =u_{n,j}^2 + \nu_{n,j}^2,
\end{equation}
which is normalized ($\sum_{j=1}^{F_m} p_{n,j}=1$). The scaling index
$\gamma_{n,j}$ for $p_{n,j}$ is defined by
\begin{equation} 
p_{n,j} \sim F_m^{-\gamma_{n,j}},
\end{equation}
In the scaling limit $m \rightarrow\infty$, according to the
multifractal theorem \cite{Kohmoto08}, the number of sites, which have
a scaling index between $\gamma$ and $\gamma+d\gamma$ is proportional
to $F_m^{f(\gamma)}$.  To distinguish the extended, critical, and
localized wave functions, only a part of $f(\gamma)$ is required. For
the extended wave functions, the maximum probability measure scales as
$\textbf{max}[p_{n,j}]\sim F_m^{-1}$; thus, we have
$\gamma_{\text{min}}=1$. For a localized wave function, $p_{n,j}$ is
finite ($\gamma=0$, $[f(0)=0]$) at some sites but on other sites
it is exponentially small ($\gamma=\infty$, $[f(\infty)=1]$); thus, we
have $\textbf{max}[p_{n,j}]\sim F_m^{0}$, or
$\gamma_{\text{min}}=0$. On the other hand, for the critical
wavefunctions, on a finite interval
$[\gamma_{\text{min}},\gamma_{\text{max}}]$, $f(\gamma)$ is a smooth
function with $0 <\gamma_{\text{min}}< 1$.  Therefore, for
distinguishing the extended, critical, and localized wave functions,
we need to calculate $\gamma_{min}$ which is defined as
$\textbf{max}[p_{n,j}]\sim F_m^{-\gamma_{min}}$. Namely,
\begin{equation}
\begin{cases}
\gamma_{\text{min}}=1, \quad \text{for an extended wave function},
\\ \gamma_{\text{min}}\neq0,1, \quad \text{for a critical wave
  function}, \\ \gamma_{\text{min}}=0, \quad \text{for a localized
  wave function.}
\end{cases}
\end{equation}
Note that here in our calculation, we plotted the average of $\gamma_{min}$ over all the eigenstates ($\overline{\gamma}_{min}$), which can be written as
\begin{equation} 
\overline{\gamma}_{min}=\frac{1}{2F_m} \sum_{n=1}^{2F_m} \gamma_{min}^n.
\end{equation}
\begin{figure}
    \centering
    \includegraphics[width=\columnwidth]{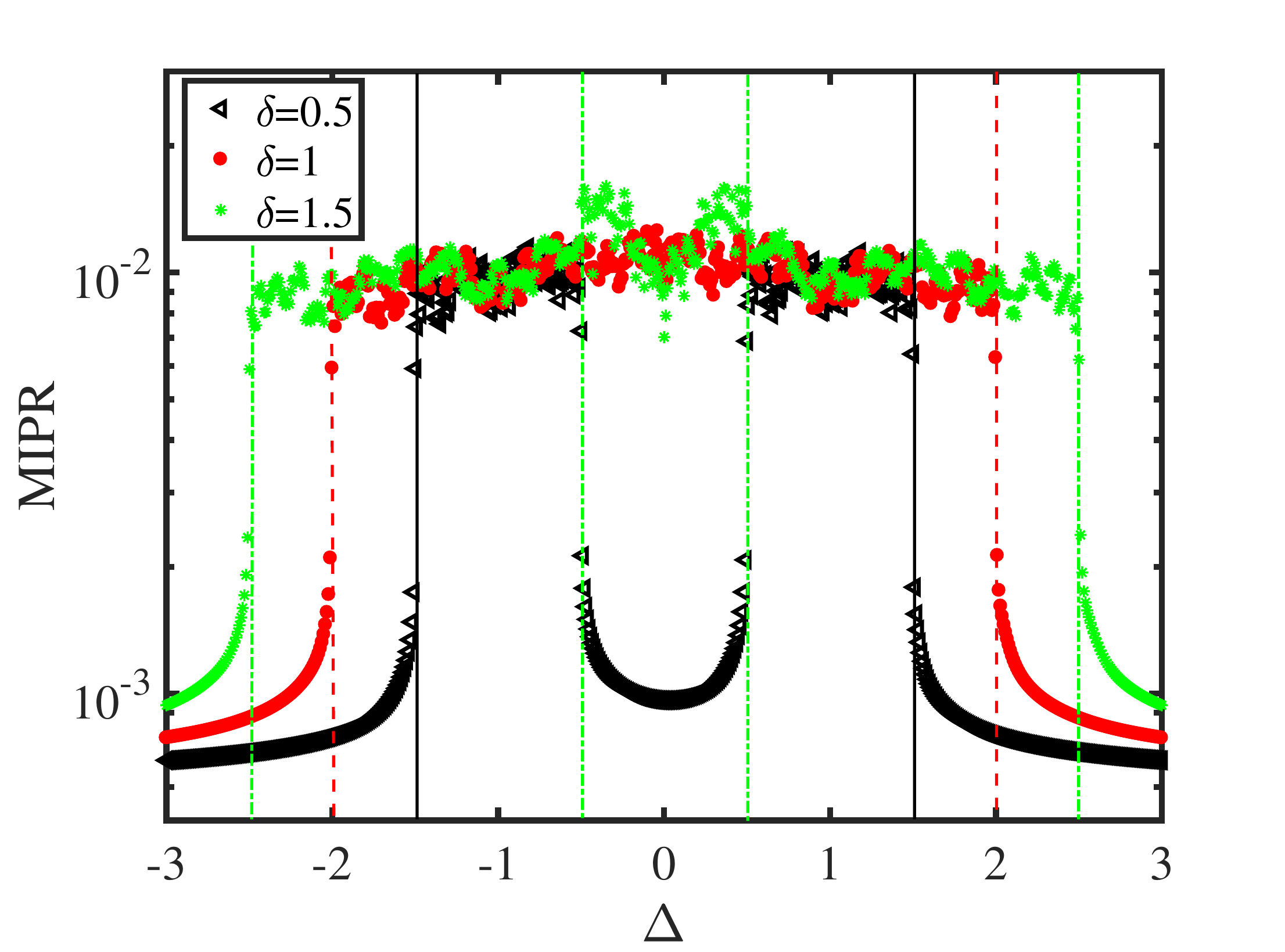}
    \caption{ MIPR as a function of $p$-wave superfluid pairing $\Delta$ for the indicated values of modulation amplitude $\delta$. The dashed
lines, dot-dashed lines, and solid lines demonstrate the abrupt changes of the MIPR at phase boundaries.}
     \label{fig:MIPR_PD}
\end{figure}
\begin{figure}
    \centering
    \includegraphics[width=\columnwidth]{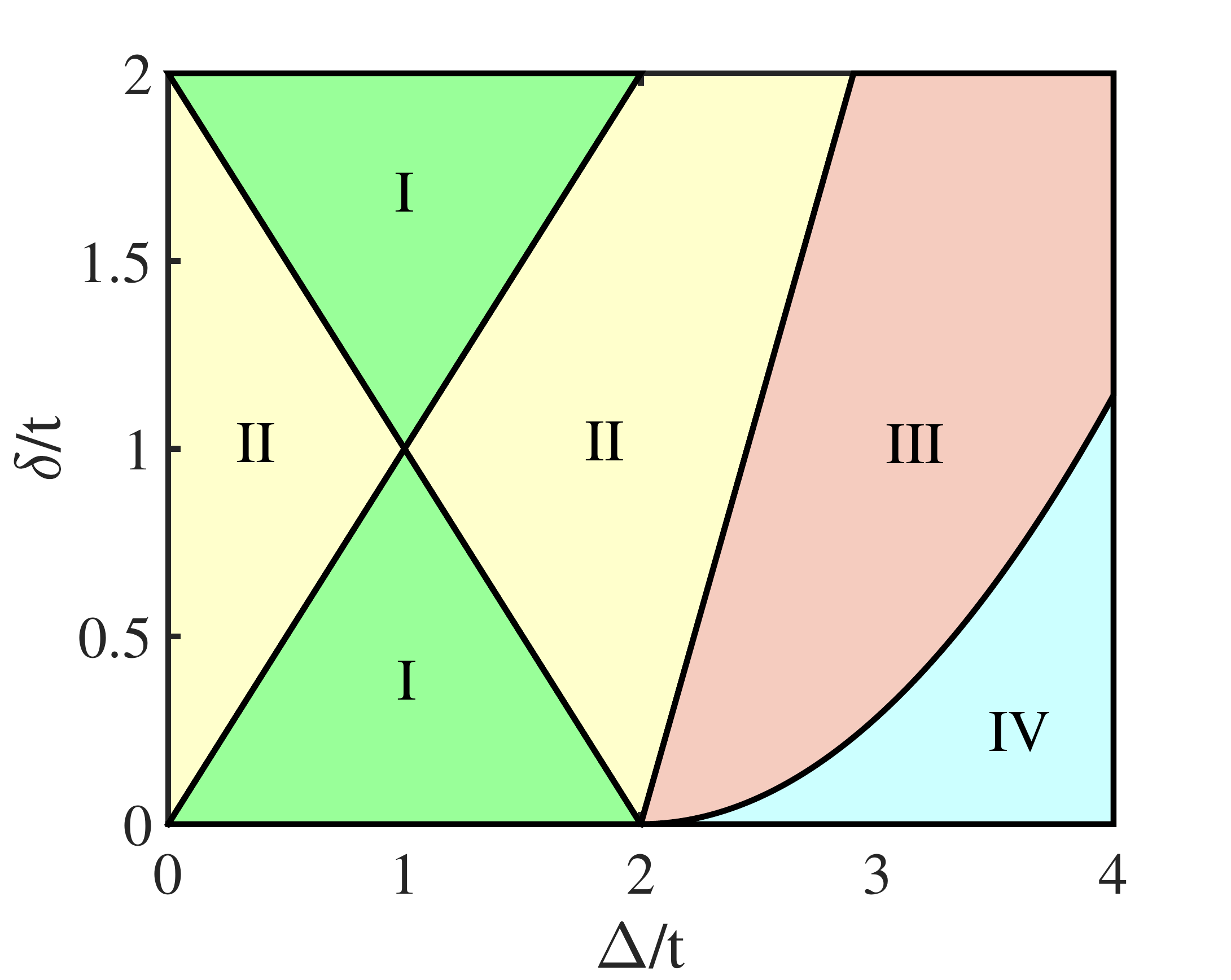}
    \caption{ Phase diagram of the off-diagonal GAAH model with the
      $p$-wave incommensurate modulation amplitude $0<\delta\leq 2$
      and the $p$-wave pairing strength $0\leq\Delta\leq4$. The
      hopping incommensurate modulation amplitude is set to $\tau = 1$
      and the phase in the incommensurate modulation is set to $k_y =
      \pi/2$ and $\varphi= \beta \pi$. The phases are (I) localized
      phase, (II) critical localized phase, (III) critical extended
      phase, and (IV) extended phase.}
     \label{fig:PD2}
\end{figure}

\begin{figure}
	\centering
	\includegraphics[width=\columnwidth]{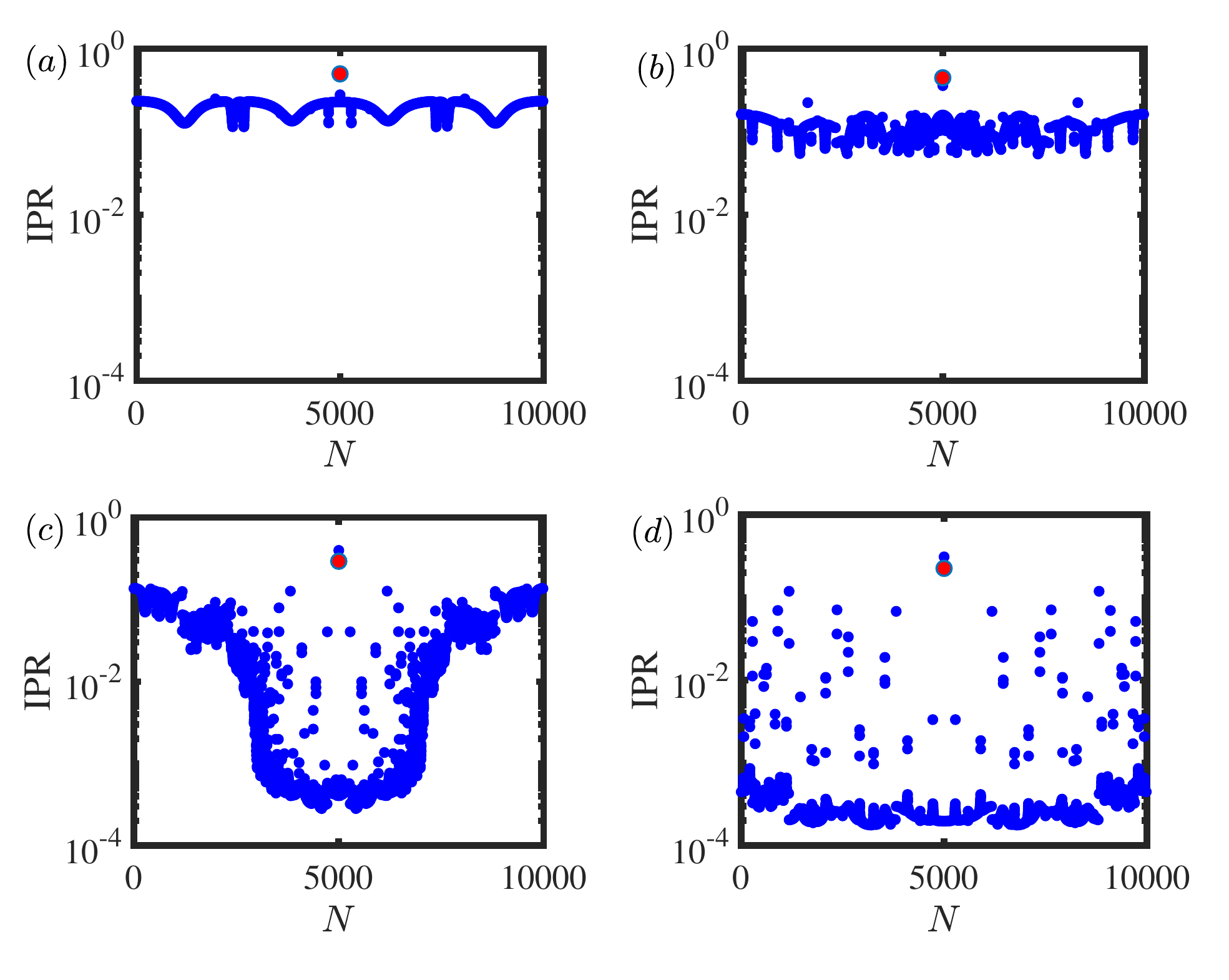}
	\caption{ The distribution of IPRs over all the eigenstates for region (a) I (localized phase), (b) II (critical localized phase), (c) III (critical extended phase), and (d) IV (extended phase) of figure \ref{fig:PD2}.}
	\label{fig:IPR_fig5}
\end{figure}

\begin{figure}
\centering
\includegraphics[width=\columnwidth]{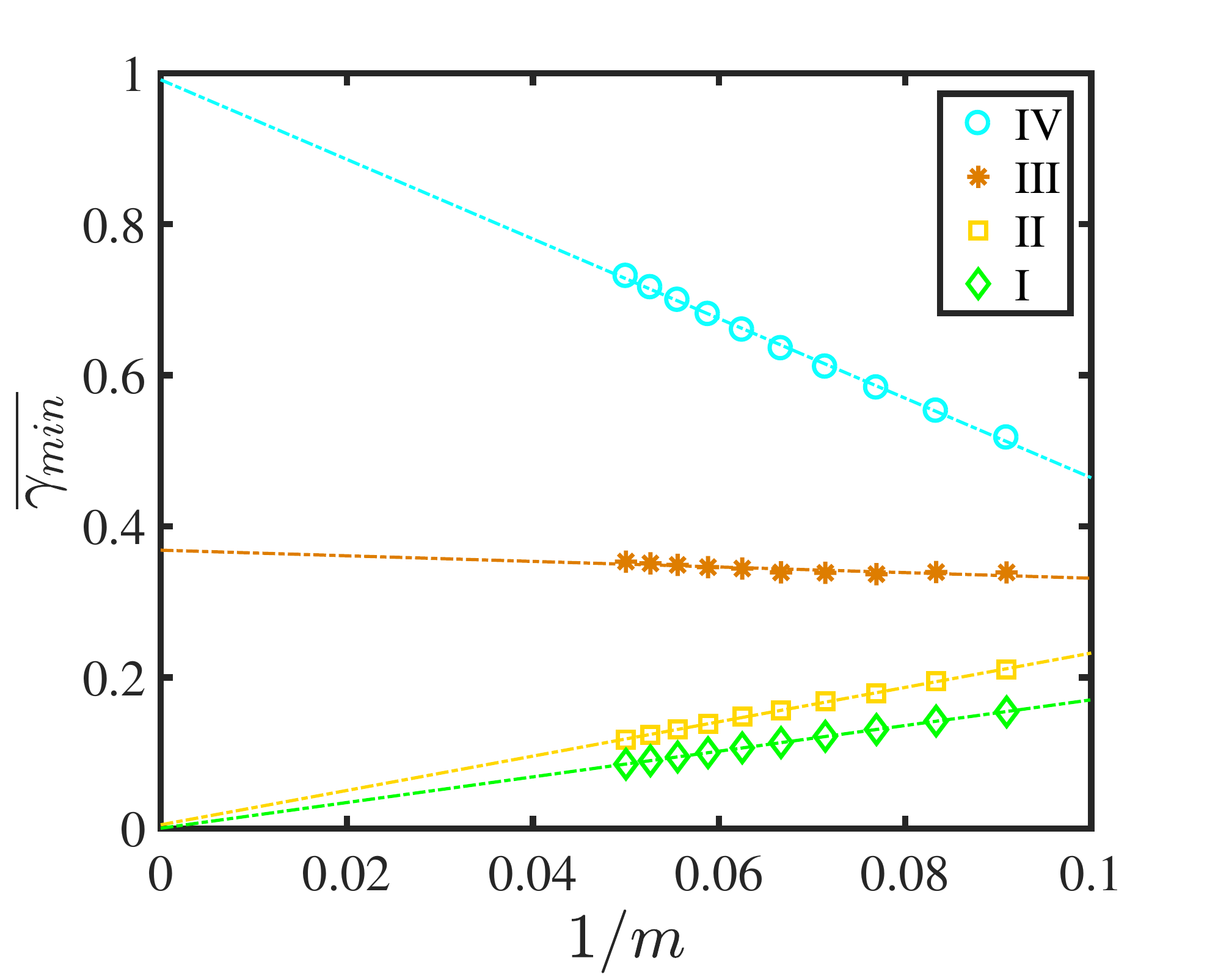}
\caption{ $\overline{\gamma_{min}}$ as function of $1/m$ for region I (localized phase), II (critical localized phase), III (critical extended phase), and IV (extended phase) of figure \ref{fig:PD2}. }
\label{fig:Multi_Fig5}
\end{figure}

\subsection{Off-diagonal GAAH model  with $p$-wave pairing }

The case $V=0$ corresponds to the off-diagonal GAAH model with
$p$-wave pairing. We calculate the phase diagram as a function of the
modulation strength of incommensurate $p$-wave pairing ($\delta$) and
the modulation strength of incommensurate hopping ($\tau$), focusing
mostly on the case $k_y=\pi/2$.  We choose $\varphi=\beta \pi$. This
off-diagonal GAAH model in the limit $\Delta=0$ and $\delta=0$
exhibits nontrivial zero-energy edge modes~\cite{Ganeshan13,Cestari16}
and in a large parameter space preserves the critical
states~\cite{Liu15}.  We find that the topological properties and
localization of this system are profoundly affected by a finite
$\delta$.  \textcolor{red}{The main new feature compared to
  Ref. \onlinecite{Liu15} is various phases with mobility edges.}  The
phase diagram based on the MIPR of the off-diagonal GAAH model with
$p$-wave pairing (without hopping modulations and site-diagonal
potential), is shown in Fig. \ref{fig:PDv0}.  The extended phase
(regions I and II), the mobility-edge phase (region III) and critical
phase (region IV) are separated by the black solid lines.  Regions I,
II, and III host two zero-energy modes as a result of nontrivial
topology which only appear for open boundary conditions. In
Fig. \ref{fig:IPR_fig2}, we show the distribution of the inverse
participation ratio (IPR) for different eigenstates. In region I
(Fig. \ref{fig:IPR_fig2}(a)), II (Fig. \ref{fig:IPR_fig2}(b)) and III
(Fig. \ref{fig:IPR_fig2}(c)), respectively, we find the zero-energy
topological edge modes (large red dots) indicating the topologically
nontrivial phase. For almost all eigenstates, the IPR distribution has
the same characteristics (around $10^{-4}$) in regions I and II, which
shows that all the eigenstates are extended.  In regions III and IV,
the value of IPR is around $10^{-2}$ which is two orders of magnitude
larger than in the extended phase. These dispersed distributions
suggests that these regions (III and IV) are critical phases. These
results confirm that regions I, II, and III are in the
nontrivial-topological phases, while the region IV is trivial
phase. Also, as shown in Fig. \ref{fig:EOBC1}, region IV is
topologically trivial and the edge modes (indicated in red color in
the figure), in the regions I and III are found to be very robust.
\textcolor{red}{For comparison, in Ref. \cite{Cestari16} the
  robustness of edge states against modulated on-site potential was
  studied, and there a critical potential strength was found beyond
  which the edge states ceased to exist.}  Here, the edge states in
occur in the case of off-diagonal disorder when $\tau=0$ and in
principle survive after the model undergoes Anderson-like
localization.

The evolution of the MIPR on a logarithmic scale at three values of
$\delta=0.5$, $1$, and $1.5$ is shown in Fig. \ref{fig:MIPR_PD}. We
find that the MIPR changes abruptly from one phase to another as a
function of $\Delta$ and $\delta$.  There are four turning points and
the change of MIPR at these points becomes sharper with increasing
system size $L$ (results not shown). Thus, in the thermodynamic limit
$L\rightarrow \infty$, a discontinuity at the turning points signals
the phase transitions among the mobility-edge, the extended, and the
critical phases.
\begin{figure}
    \centering
    \includegraphics[width=\columnwidth]{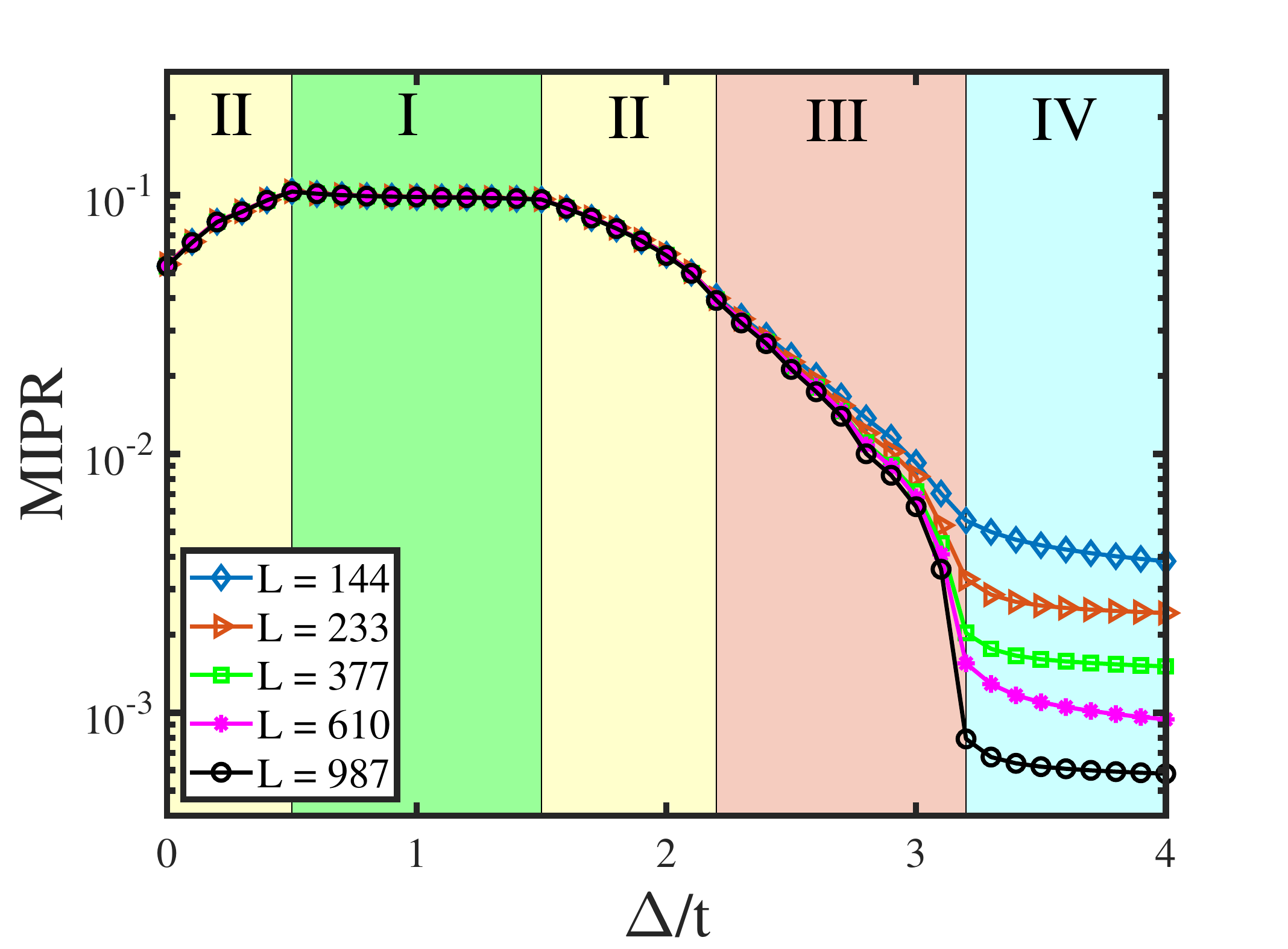}
    \includegraphics[width=\columnwidth]{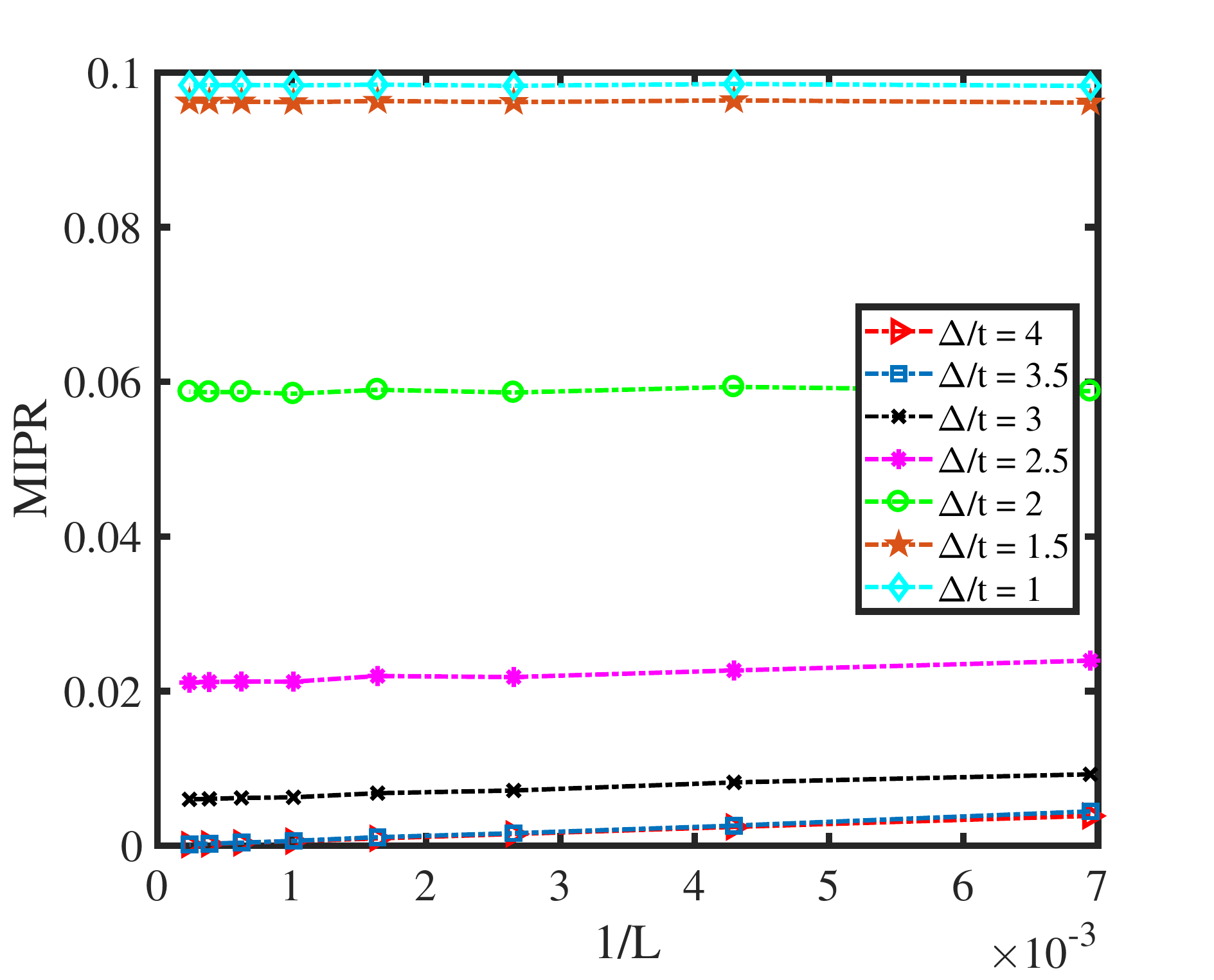}
    \caption{ Upper panel: MIPR (log-scale) as a function of $p$-wave superfluid pairing $\Delta/t$ for different chain lengths $L$ for $\tau=1$ and $\delta=0.5$. Bottom panel: MIPR with the inverse system size $1/L$. For the extended phase, MIPR tends to zero as $L$ increases.}
     \label{fig:MIPR2}
\end{figure}
We have also considered the dependence of the phase diagram on nonzero
$\tau$. By calculating the MIPR, we find that the localization
properties of this model are significantly affected by turning on the
off-diagonal hopping modulation of $\tau$. For $\tau=1$, the results
are summarized in the phase diagram shown in Fig. \ref{fig:PD2}. There
are four distinct phases, localized phase (I), critical localized
phase (II), critical extended phase (III), and extended phase (IV),
separated by solid black lines.  In Fig. \ref{fig:IPR_fig5}, we show
examples of the distribution of IPR over different eigenstates for
localized phase (I), critical localized phase (II), critical extended
phase (III), and extended phase (IV) of Fig. \ref{fig:PD2}.
Topological edge states are found in all phases.  An interesting
situation is depicted in Fig. \ref{fig:IPR_fig5} (c): the MIPR
indicates the simultaneous presence of localized and extended states,
as in a mobility edge phase, but here the boundary is smeared between
the two.  As the eigenenergy increases in Fig.  \ref{fig:IPR_fig5}(c),
the IPR smoothly changes from a typical value for the localized stets
around $10^{-1}$ to a typical value for the extended states
$10^{-4}$. The smooth changes of the IPR suggest that there exist the
semi-mobility edge in the energy spectrum.  Also, for these selected
phases in Fig. \ref{fig:Multi_Fig5}, we plotted
$\overline{\gamma}_{min}$ as a function of $1/m$. For the localized
phase, $\overline{\gamma}_{min}$ extrapolates to $0$ and but critical
localized phase, $\overline{\gamma}_{min}$ vanish to
zero. Furthermore, for critical extended phase
$\overline{\gamma}_{min}$ extrapolates to $0.38$, but for extended
phase $\overline{\gamma}_{min}$ extrapolates to $1$. These results
also confirm our phase diagram in Fig. \ref{fig:PD2}.

Fig. \ref{fig:MIPR2} shows the MIPR of the model as a function of
$\Delta$ with $\tau=1$ and $\delta = 0.5$ for different system
sizes. We have checked that with increasing $L$, the system is in the
localized region (I) for $\delta-\tau+1<\Delta<\tau-\delta+1$. Also,
for the extended phase (III) the MIPR is finite and depend on the
system size ($L$). In this phase, the MIPR satisfies the finite size
scaling (FSS) form, $\textmd{MIPR}=b L^{-\eta}$. At the
$\Delta/t=3.5$, $\eta=0.63$. For this phase, with the increase of $L$,
MIPR tends to zero. The MIPR among localized, critically localized,
critical extended, and extended phases satisfies
\begin{equation}\label{Order}
  \textmd{MIPR}_E< \textmd{MIPR}_{CE}<\textmd{MIPR}_{CL}<\textmd{MIPR}_L
\end{equation}
We verified this expression by checking the FSS in the whole phase
diagram (results not shown).

\begin{figure}
    \centering
    \includegraphics[width=\columnwidth]{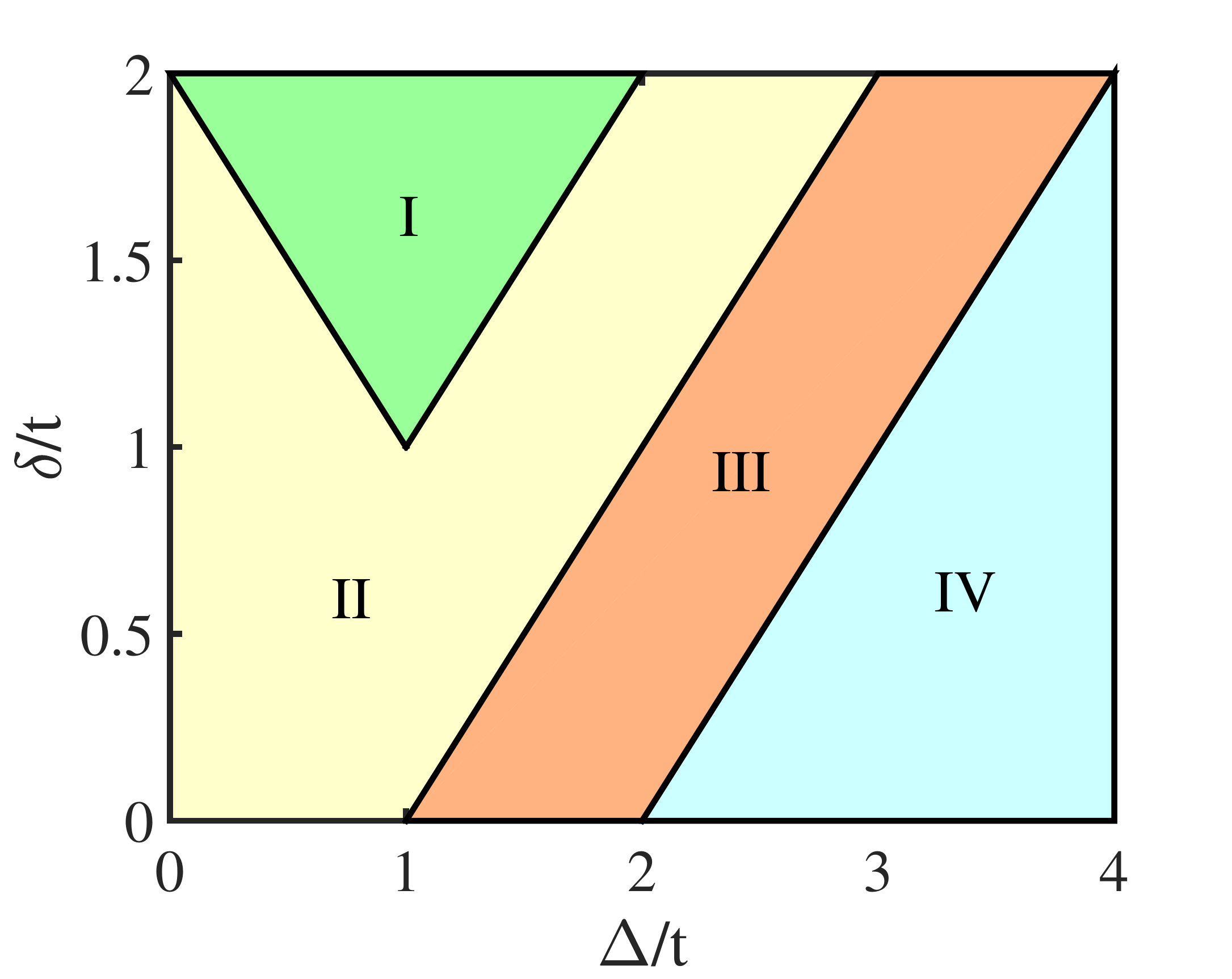}
    \caption{ Phase diagram of the generic GAAH model with the
      $p$-wave incommensurate modulation amplitude $0<\delta\leq 2$
      and the $p$-wave pairing strength $0\leq\Delta\leq4$. The
      hopping incommensurate modulation amplitude is set to $\tau =
      1$, the on-site potential incommensurate modulation amplitude is
      set to $V/t = 1$, and the phase in the incommensurate modulation
      is set to $k_y = \pi/2$ and $\varphi= \beta \pi$. The phases are
      (I) localized phase, (II) critical localized phase, (III)
      critical extended phases, (IV) extended phases.}
     \label{fig:PDV1}
\end{figure}

\begin{figure}
	\centering
	\includegraphics[width=\columnwidth]{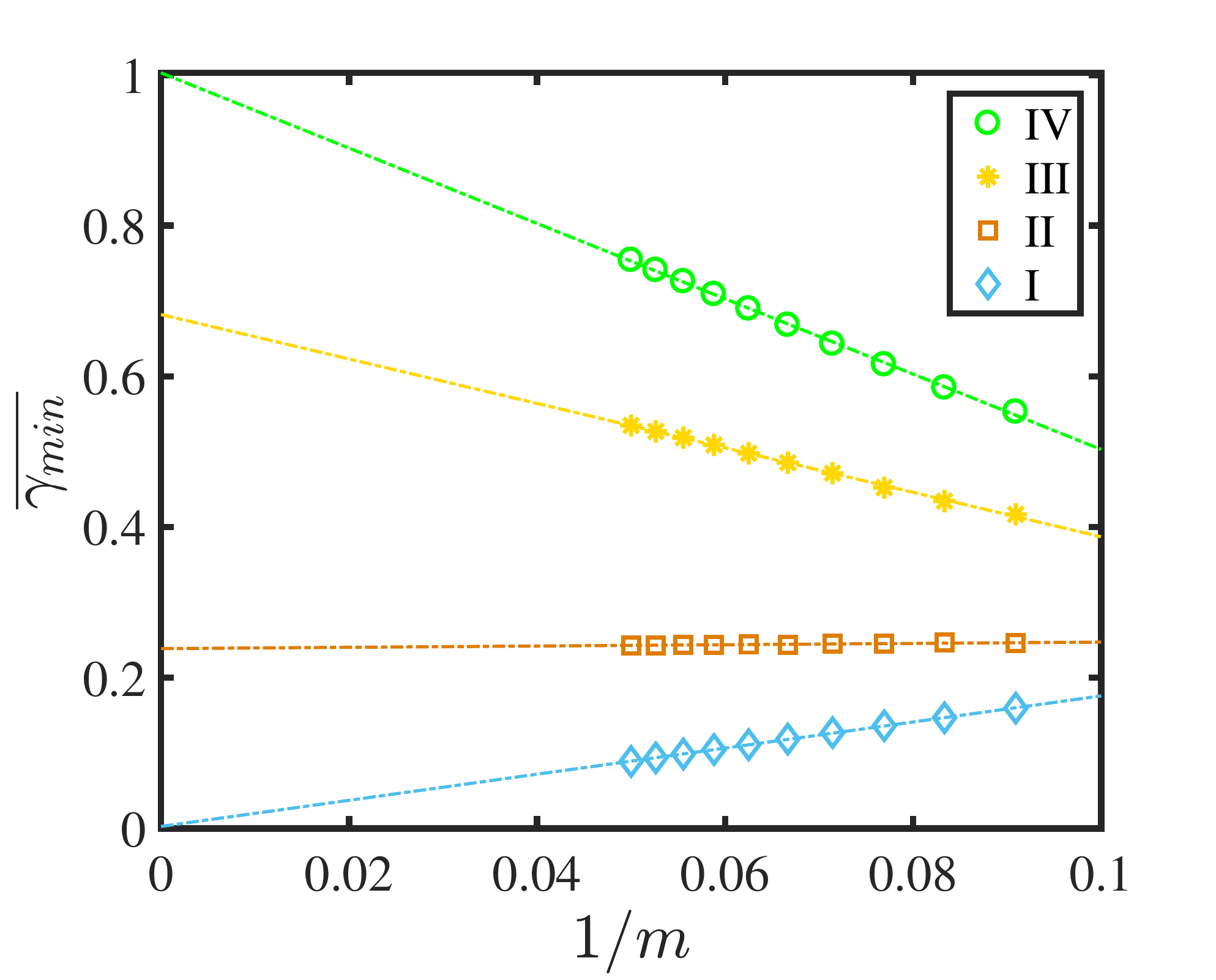}
	\caption{ $\overline{\gamma_{min}}$ as function of $1/m$ for
          region I (localized phase), II (critical localized phase),
          III (critical extended phase), and IV (extended phase) of
          figure \ref{fig:PDV1}. }
	\label{fig:Multi_Fig7}
\end{figure}

\begin{figure}
	\centering
	\includegraphics[width=\columnwidth]{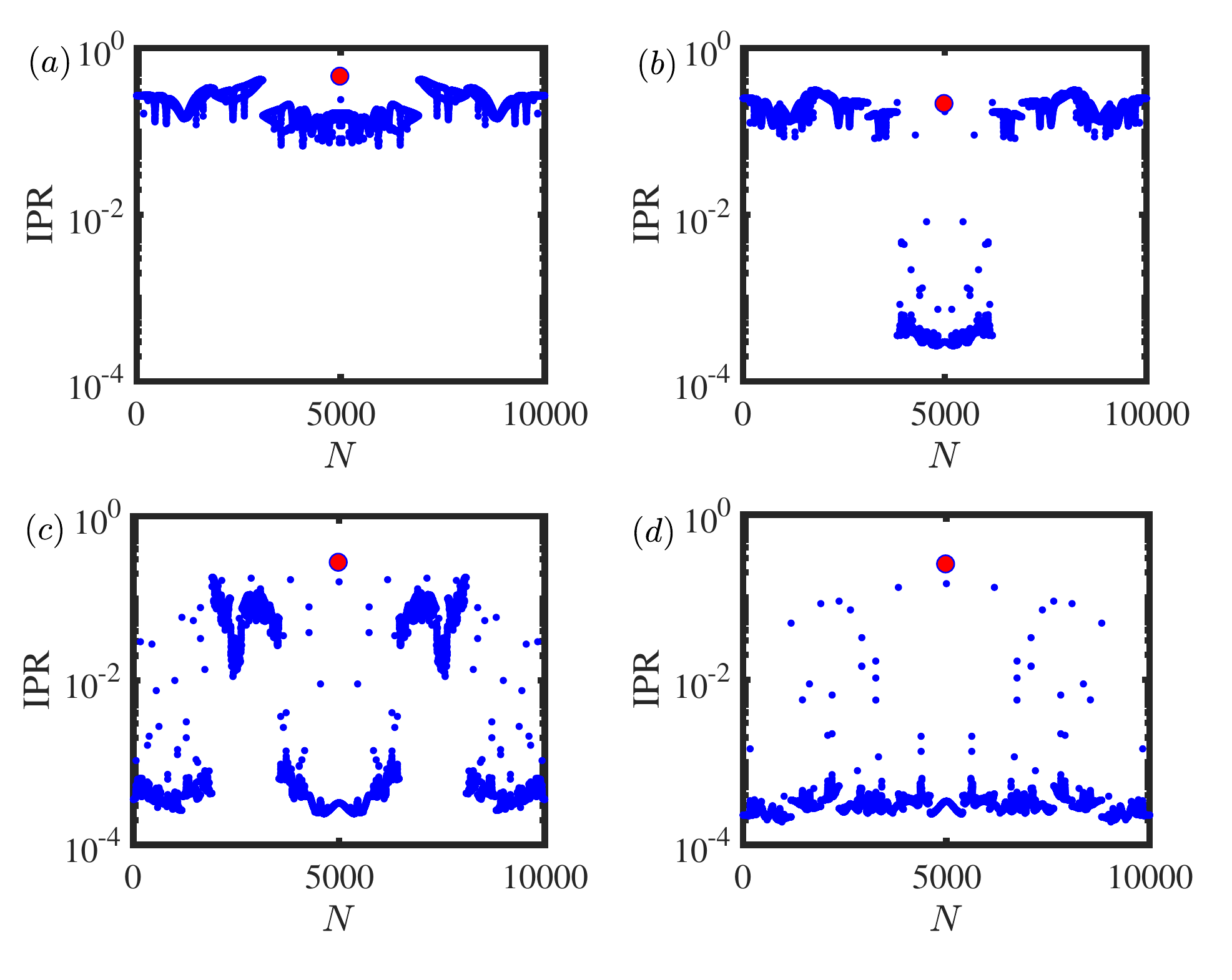}
	\caption{ The distribution of IPRs over all the eigenstates
          for (a) region I (localized phase), (b) II (critical
          localized phase), (c) III (critical extended phase), and (d)
          IV (extended phase) of figure \ref{fig:PDV1}. }
	\label{fig:IPR_fig7}
\end{figure}

\begin{figure*}[ht]
\includegraphics[width=\linewidth]{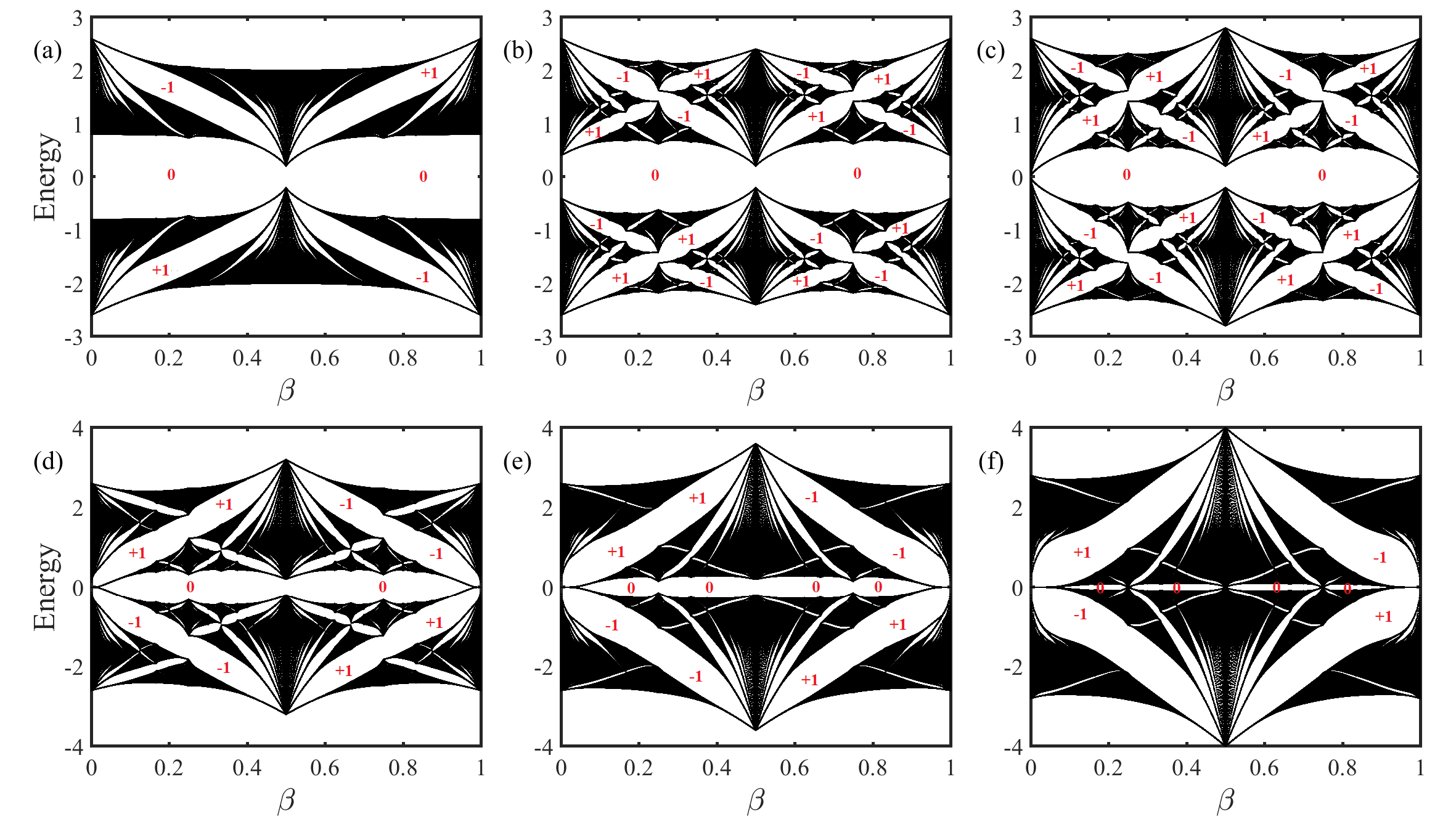}
\caption{ Hofstadter butterfly: energy spectrum as a function of
  magnetic flux per plaquette $\beta$ in Aubry-Andr{\'e} lattice with
  $V=0$, $\tau=0.3$, $\Delta=0.4$, and (a) $\delta=0$, (b)
  $\delta=0.2$, (c) $\delta=0.4$, (d) $\delta=0.6$, (e) $\delta=0.8$,
  (f) $\delta=1$.}\label{fig:bf1}
\end{figure*}

\subsection{Generic GAAH model  with $p$-wave pairing }

We also investigate the generic $p$-wave pairing GAAH model with
modulated on-site potentials, modulated off-diagonal hopping terms and
modulated $p$-wave pairing terms. In this section, we explore the
influence of the modulated on-site potential on the phase transition.
It is clear that varying $V$ changes the phase diagram. So due to the
modulation in the $p$-wave pairing term, we find that the system has a
stronger tendency to become extended as a function of $\Delta/t$ when
the disordered on-site potential ($V$) is varied. In
Fig. \ref{fig:PDV1}, the phase diagram of the generic GAAH model for
the case $V/t = 1$ is shown. The phase boundary separating the
localized phase, critical localized phase, critical extended phases,
and extended phases vary rapidly with the SC pairing. We also
performed a FSS analysis for this phase diagram (results not
shown). When $V/t = 1$, it is clear that for $\delta/t<1$, the
localized phase disappears and the critical localized phase increases
sharply.  We also focus on the distribution of IPR with different
eigenstates and multifractal analysis~\cite{Hiramoto89} for this
case. Example of determining $\overline{\gamma}_{min}$ as a function
of $1/m$ in phases I, II, III, and IV of Fig. \ref{fig:PDV1} is shown
in Fig.  \ref{fig:Multi_Fig7}. For this phase diagram,
$\overline{\gamma}_{min}$ extrapolates to $0$ if we are in I-phase (in
this case, the multifractal analysis of the wave function shows that
all wave functions are in localized states) and to $1$ if are in
IV-phase (in this case, the multifractal analysis of wave function
shows that all of the wave functions are in extended state).  The
distribution of IPR in the eigenstates, shown in
Fig. \ref{fig:IPR_fig7}, indicates that almost all the eigenstates IPR
are close to each other in phase I (being around $10^{-1}$) and phase
IV (being around $10^{-4}$). For phases II and III, the multifractal
analysis of wave function is shown in Fig. \ref{fig:Multi_Fig7} (b)
and Fig. \ref{fig:Multi_Fig7} (c). For the critical localized phase
(II), $\overline{\gamma}_{min}$ extrapolates to $0.245$ and for the
critical extended phase (III,) $\overline{\gamma}_{min}$ extrapolates
to $0.685$.  Again, as in Fig. \ref{fig:Multi_Fig5}, all phases
exhibit topological edge modes.  As the eigenenergy increases in Fig.
\ref{fig:IPR_fig7}(b) for the critical localized state, the IPR
suddenly jumps from a typical value for the localized states around
$10^{-1}$ to a typical value for the extended states $10^{-4}$, but
for the critical extended state this happens twice (see Fig.
\ref{fig:IPR_fig7}(c)).  \textcolor{red}{Models of which we are
  aware~\cite{Sarma10,Liu15} show one mobility edge jump.}  Recall
that in Fig. \ref{fig:Multi_Fig5} it was also the critical extended
state which showed unusual behavior, the smeared mobility edge.

In summary we find that due to the modulation in the SC pairing the
incommensurate generic GAAH model with $p$-wave pairing delocalizes
easier when varying the disordered on-site potential, when
$\delta<V$. We find that the topological properties of the generic
GAAH model with $p$-wave pairing are significantly affected by turning
on the modulated on-site potential and modulated $p$-wave SC pairing.

\begin{figure}[ht]
\includegraphics[width=\linewidth]{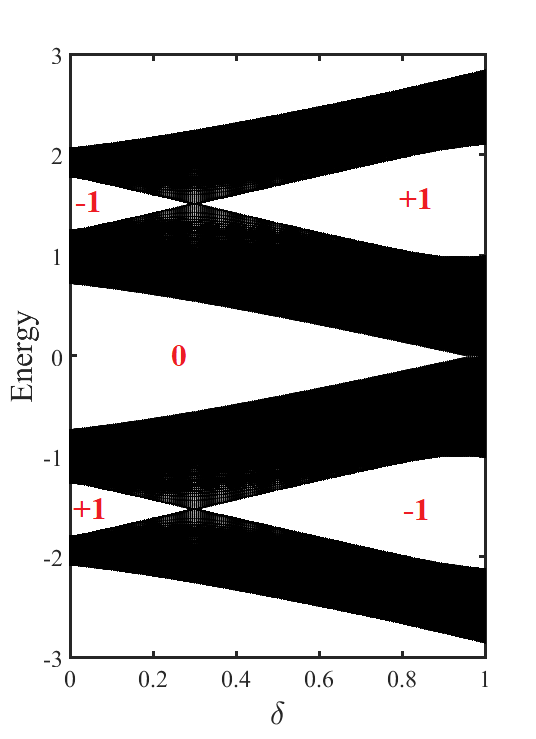}
\caption{The evolution of the energy bands for $\beta=0.25$,  $V=0$, $\tau=0.3$, and $\Delta=0.4$  as function of $\delta$ shown in Fig. (\ref{fig:bf1}). Gaps are labeled with their
Chern numbers.}\label{fig:HF1}
\end{figure}

\begin{figure*}[ht]
\includegraphics[width=\linewidth]{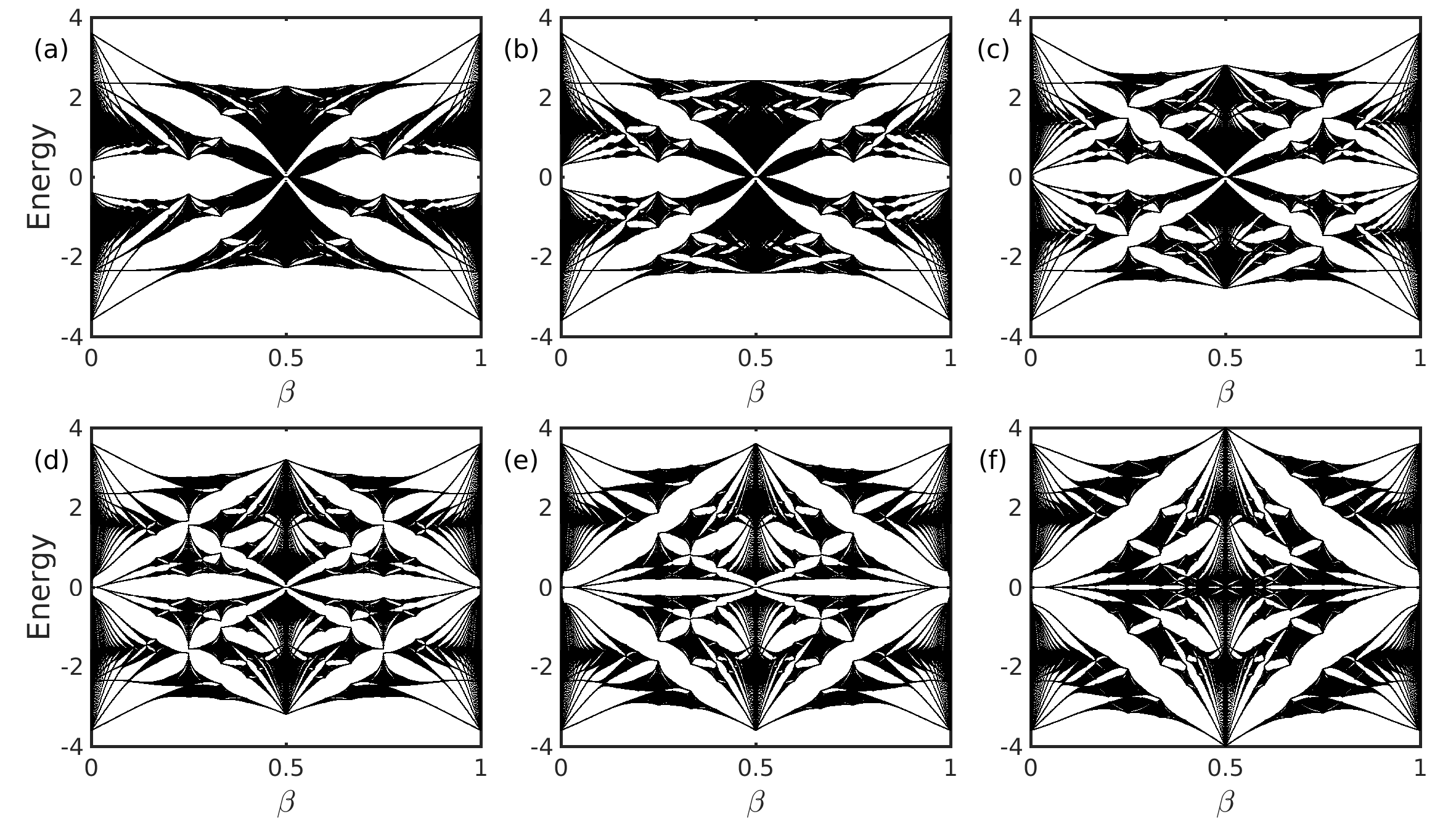}
\caption{ Hofstadter butterfly: energy spectrum as a function of magnetic flux per plaquette $\beta$ in Aubry-Andr{\'e} lattice with $V=1$, $\tau=0.3$, $\Delta=0.4$, and (a) $\delta=0$, (b) $\delta=0.2$, (c) $\delta=0.4$, (d) $\delta=0.6$, (e) $\delta=0.8$, (f) $\delta=1$.}\label{fig:bf2}
\end{figure*}

\begin{figure}[ht]
\includegraphics[width=\linewidth]{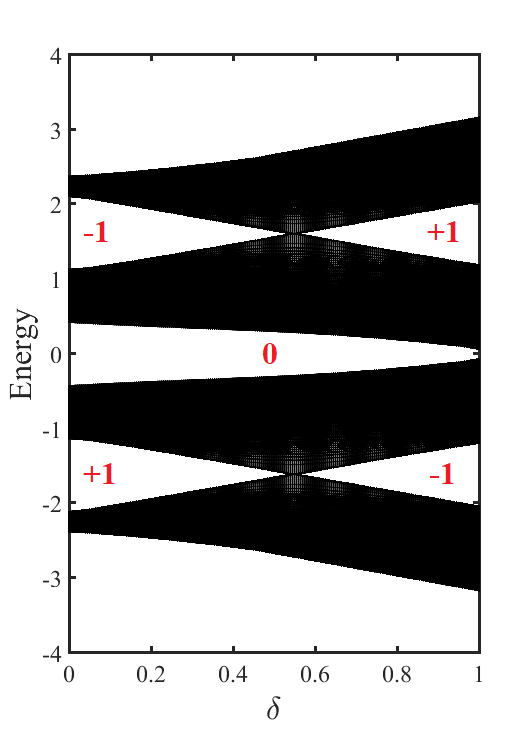}
\caption{The evolution of the energy bands for $\beta=0.25$,  $V=1$, $\tau=0.3$, and $\Delta=0.4$  as function of $\delta$ shown in Fig. (\ref{fig:bf2}). Gaps are labeled with their
Chern numbers.}\label{fig:HF2}
\end{figure}
\section{Commensurate modulation}
\label{Sec_C}
When $\beta$ is rational, the lattice is commensurate. It is
known that in the commensurate case \cite{Zeng16}, the system will not
undergo a localization-delocalization transition as in the
incommensurate case. When $\beta = 1/2$, both Kitaev-like and
Su-Schrieffer-Heeger-like (SSH-like) models are included in this GAAH
model with $p$-wave SC pairing for commensurate modulations. The
Hamiltonian of Eq. (\ref{GAA-PW}) is reduced to the SSH model for $V
=\Delta =\delta=0$ and to the Kitaev model for $\tau=\delta= 0$. In
order to determine the different phase boundaries and characterize the
topological phases, we need to calculate the effect of modulated SC
pairing on the topological properties of the system. In the following,
we characterize the topological nature of the modulated SC pairing by
calculating the evolution of Chern numbers~\cite{Avron83} for the
major gaps of the spectrum.

\subsection{Chern numbers}
\begin{figure}[ht]
\includegraphics[width=\linewidth]{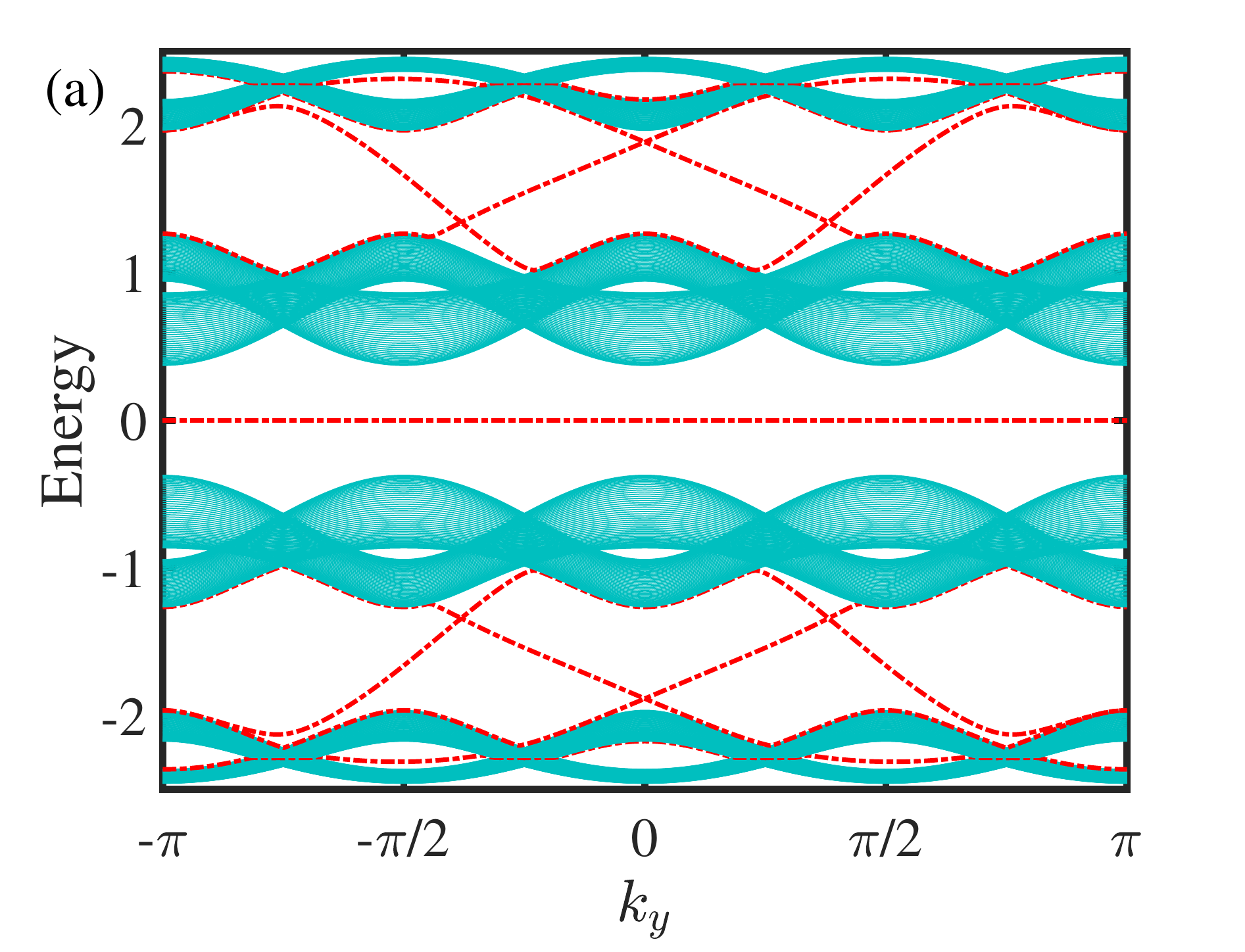}\\
\includegraphics[width=\linewidth]{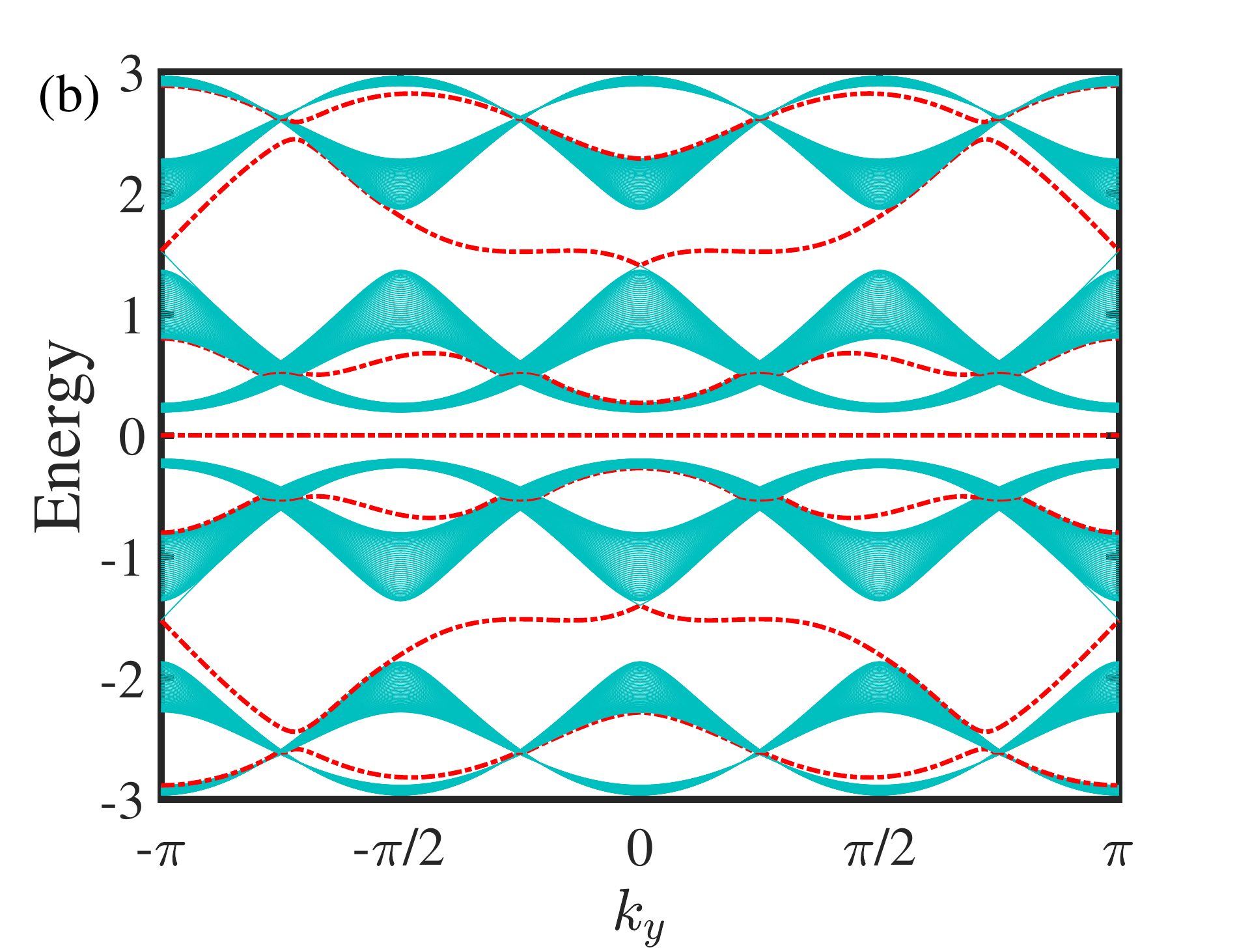}
\caption{The evolution of the energy bands for $\beta=0.25$,  $V=1$, $\tau=0.3$, and $\Delta=0.4$  as function of $\delta$ shown in Fig. (\ref{fig:bf2}). Gaps are labeled with their
Chern numbers.}\label{fig:HF2}
\end{figure}
Chern numbers can be calculated from the density with respect to 
changes in the magnetic field using the St\v{r}eda formula
\cite{Streda82,Umucallar08}.  In lattice systems, the Chern number can
be written as
\begin{equation}\label{SF}
  C=    \frac{\partial \bar{n}(\beta)}{\partial \beta}
\end{equation}
where $\bar{n}$ is the number of levels below the Fermi level.  This
formula is valid when the chemical potential lies in a
gap~\cite{Streda82}. Formally, the evaluation of the Chern numbers can
be calculated by $k$-space integration of the Berry curvature over the
Brillouin zone. In Fig. \ref{fig:bf1} we present the Hofstadter
butterfly for the case $V/t=0$, $\lambda=0.3$, $\Delta=0.4$ and
various values of $\delta$. The spectrum for $\delta=0$ is clearly
formed by two triangular-lattice Hofstadter butterflies separated by a
large gap. The Hofstadter butterfly separation is controlled by the SC
parameter, $\delta$ (see Fig. \ref{fig:bf1}). As illustrated in this
figure, by changing $\delta$, the upper (lower) butterfly approaches
the square regime and then the honeycomb regime, which has different
topology (in other words at the $\beta=0.2$ for upper lattice the
Chern number changes from $-1$ to $1$).  Also, we see that as $\delta$
is increased, the gaps at zero-energy are sufficiently small and have
not completely closed so that the overall the butterfly shape is
maintained. It is possible to understand this surprising result by
considering the evolution of the energy bands at the points at which
the main gap closes and reopens. For $V/t=0$, $\lambda=0.3$,
$\Delta=0.4$, and $\beta=0.25$, these evolutions are demonstrated in
Fig. \ref{fig:HF1}. When the energy gap closes and reopens through
this evolution, Chern numbers will change.  For this case, we find two
different regions corresponding to different Chern numbers, $\pm 1$.
An important observation is that the Chern number is sensitively to
the $\delta$ parameter. As can be seen in Fig.  \ref{fig:HF1}, the
whole area is covered by the triangular lattice, while the square
lattice is confined to the point at gap closure, $\delta=0.3$.
 
For the second case we consider $V\neq0$, when the system has
sublattice asymmetry. In this case, we can see that when $V\neq0$, the
evolution of the energy spectrum between the triangular lattice and
the square lattice remain unchanged (see Fig. \ref{fig:bf2}). More
interestingly, in this case, parts of the energy modes in the
topologically nontrivial regime split into two energy modes which can
be taken as two 1D Majorana chains coupled by the on-site
potential. As an example in Fig. \ref{fig:HF2}, we display the
evolution of bands from the triangular to the square and then
triangular with different topology. As is shown in this figure, when
$V$ becomes nonzero and gets stronger, the critical value of the phase
transition $\delta_c$ is increased.

\textcolor{red}{
\section{CONCLUSION}
\label{Con}
In this paper, we studied a GAAH model with modulated $p$-wave
SC-pairing, both in the commensurate and incommensurate cases.  We
mapped derived the 2D magnetic analog of the model, have shown that in
the bipartite commensurate case four different topological excitations
are possible, mapped the phase diagram of the model via studying the
localization characteristics, and studied the evolution of its
Hofstadter butterflies.  The phase diagram is remarkably rich,
exhibiting localized, extended, and critical phases, as well as
topological edge states, which can occur in extended or critical
cases.  Several new phases were revealed, unique to the model with
modulated $p$-wave pairing: in the critical extended phase when
incommensurate $p$-wave pairing and hopping is turned on, the change
change in inverse participation ratios separating extended and
localized regions is smeared, rather than sharp, as it
happens~\cite{Liu18} in the extended GAAH model without $p$-wave
pairing modulation.  When, in addition, the on-site potential is
modulated the jumps between extended and localized regions are again
sharp, but increase in number.  For the commensurate case the
modulated SC amplitude results in Hofstadter butterfly plots showing a
transition from the rectangular to the square lattice as the parameter
$\delta$ varies.  }
\section*{Acknowledgments}
B.H. was supported by the National Research, Development and
Innovation Fund of Hungary within the Quantum Technology National
Excellence Program (Project Nr.  2017-1.2.1-NKP-2017-00001).  B.T. and
M.Y. were supported by the he Scientific and Technological Research
Council of Turkey (TUBITAK) under Grant No. 117F125. B.T. is also
supported by TUBA.

\end{document}